\documentclass[11pt,a4paper]{article}

\usepackage[utf8]{inputenc}
\usepackage[T1]{fontenc}
\usepackage{lmodern}
\usepackage[english]{babel}

\usepackage{geometry}
\usepackage{microtype}
\usepackage{amsmath}
\usepackage{amssymb}
\usepackage{graphicx}
\usepackage{booktabs}
\usepackage{longtable}
\usepackage{array}
\usepackage{xcolor}
\usepackage{listings}
\usepackage{textcomp}
\usepackage[super,sort&compress]{natbib}
\usepackage[hidelinks]{hyperref}
\usepackage{cleveref}
\usepackage{authblk}

\usepackage[section]{placeins}
\usepackage{etoolbox}
\preto\subsection{\FloatBarrier}

\lstdefinestyle{shellstyle}{
  basicstyle=\ttfamily\footnotesize,
  breaklines=true,
  breakatwhitespace=true,
  showstringspaces=false,
  frame=single,
  framesep=3pt,
  framerule=0.4pt,
  rulecolor=\color{black!20},
  backgroundcolor=\color{black!2},
  xleftmargin=1em,
  xrightmargin=1em,
  columns=fullflexible,
  keepspaces=true,
  language=bash,
  morekeywords={ai2-kit,omb,sbatch,mpirun,xargs,ls,bash},
  commentstyle=\color{black!50},
}
\lstdefinestyle{pythonstyle}{
  basicstyle=\ttfamily\footnotesize,
  breaklines=true,
  breakatwhitespace=true,
  showstringspaces=false,
  frame=single,
  framesep=3pt,
  framerule=0.4pt,
  rulecolor=\color{black!20},
  backgroundcolor=\color{black!2},
  xleftmargin=1em,
  xrightmargin=1em,
  columns=fullflexible,
  keepspaces=true,
  language=Python,
  commentstyle=\color{black!50},
}
\newcommand{\AItwo}{AI\textsuperscript{2}}

\title{Ai2-Kit: Streamlining AI-Accelerated Ab Initio Workflows for Complex Chemical Systems}

\author[1,2,5]{Sheng Bi}
\author[5]{Wei-Hong Xu}
\author[1]{Yong-Bin Zhuang}
\author[1]{Jia-Xin Zhu}
\author[1]{Jiang-Peng Qiu}
\author[1]{Yu-Hang Tang}
\author[1]{Xiang-Long Du}
\author[1]{Qi You}
\author[1,5]{Yun-Pei Liu}
\author[1]{Fu-Qiang Gong}
\author[1]{Yu-Xin Guo}
\author[1]{Yi-Ze Wang}
\author[1]{Cheng-Xuan Wang}
\author[1]{Zi-Heng Gong}
\author[1]{Zi-Qiang Chen}
\author[3]{Chang Liu}
\author[1]{Siyuan Han}
\author[1]{Jian Gu}
\author[1]{Jia-Xin Li}
\author[1]{Yi-Ming Chen}
\author[1]{Lin Huang}
\author[1]{Si-Jie Chen}
\author[1]{Bo-Ying Huang}
\author[1]{Jie-Zhen Xia}
\author[3]{Fan-Jie Xu}
\author[1]{Su-Yang Zhong}
\author[3]{Peng-Wei Xu}
\author[1]{Jun-Yi Wang}
\author[1]{Xing-Yun Xie}
\author[3]{Yu-Lei Gong}
\author[1]{Yan-Yi Su}
\author[1]{Yue Liu}
\author[1]{Rui-Hao Bi}
\author[1]{Lang Li}
\author[1]{Fei-Teng Wang}
\author[5]{Jing-Xiang Zou}
\author[1]{Mei Jia}
\author[1]{Jie-Qiong Li}
\author[1]{Min Lin}
\author[1]{Qi-Yuan Fan}
\author[1]{Juan-Juan Sun}
\author[1]{Jia-Bo Le}
\author[1]{Zixuan Wei}
\author[1]{Jin-Yuan Hu}
\author[1,5]{Meng-Lei Jia}
\author[1]{Yan Sun}
\author[4,5,*]{Xiao-Hui Yang}
\author[1,3,5,*]{Fujie Tang}
\author[1,5,*]{Feng Wang}
\author[1,3,5,*]{Jun Cheng}

\affil[1]{State Key Laboratory of Physical Chemistry of Solid Surfaces, iChEM, College of Chemistry and Chemical Engineering, Xiamen University, Xiamen, China}
\affil[2]{College of Materials, Xiamen University, Xiamen, China}
\affil[3]{Institute of Artificial Intelligence, Xiamen University, Xiamen, China.}
\affil[4]{National Engineering Research Center of Chemicals for Electronic Manufacturing (Reconstruction), College of Chemistry and Chemical Engineering, Xiamen University, Xiamen, China}
\affil[5]{Laboratory of AI for Electrochemistry (AI4EC), IKKEM, Xiamen, China}

\affil[*]{Corresponding authors. E-mail:
\href{mailto:xiaohuiyang@xmu.edu.cn}{xiaohuiyang@xmu.edu.cn},
\href{mailto:tangfujie@xmu.edu.cn}{tangfujie@xmu.edu.cn},
\href{mailto:fengwang@xmu.edu.cn}{fengwang@xmu.edu.cn},
\href{mailto:chengjun@xmu.edu.cn}{chengjun@xmu.edu.cn}}

\date{}

\begin{document}
\maketitle

\begin{abstract}
Molecular simulations of complex chemical systems, such as catalysis, electrochemistry,
and energy storage, often need to capture the interplay of effects such as electronic
structure, finite-temperature fluctuations, and electric-field response. Such complexity
is difficult to address with traditional ab initio calculations, which are limited by
the time and length scales they can reach. AI-accelerated ab initio
(\AItwo{}) methods use machine learning potentials trained on first-principles data to
replace expensive electronic-structure calculations, extending ab initio accuracy to
these regimes, but their routine application requires reliable workflows that connect
first-principles calculations, model training, molecular dynamics, enhanced sampling,
trajectory analysis, and HPC orchestration. Here we present ai2-kit, a software toolkit for developing
accessible, reproducible, and extensible \AItwo{} workflows. ai2-kit provides
high-semantic-density command-line interfaces and Python APIs for structure and dataset
conversion, batch task generation, active-learning screening, job orchestration, and
workflow recovery. We demonstrate ai2-kit in four representative applications:
active-learning-based machine learning potential construction, free-energy perturbation
for redox and acid--base processes, electrochemical machine learning potentials for
electrified interfaces, and spectroscopies from machine learning molecular dynamics. ai2-kit
also provides AI-agent skills that help users adapt these use cases into customized
workflows for their own chemical systems and computational software stacks. Together,
ai2-kit helps turn \AItwo{} methods from bespoke computational protocols into reusable
and extensible workflows for complex chemical systems, from model construction to property
prediction.
\end{abstract}

\section{Introduction}
\label{sec:intro}

Complex chemical systems are characterized by the strong coupling between atomic
structure, electronic response, and the surrounding environment across multiple time and
length scales.~\citep{truhlar2008complex,gastegger2021solvent} Such systems are central to many frontier areas in chemistry and materials science,
including catalysis, electrochemistry, energy storage and conversion, and the design of
functional materials.~\citep{goldsmith2018catalysis,pt2024doublelayer,wang2024redox,cheng2021water} In such systems, chemical behavior cannot be fully understood from a single
optimized structure alone; it emerges from ensembles of configurations shaped by finite
temperature, solvation, interfaces, electric fields, and chemical reactions. This complexity
is reflected in the processes and observables used to characterize such systems, including
proton and ion transfer in solution,~\citep{wang2022electrolyte,wang2023redox,wang2024redox,sno2proton2024} the structure and response of electric double
layers,~\citep{kornyshev2007double,bazant2011overscreening,pt2024doublelayer} and the vibrational spectra of confined or interfacial liquids.~\citep{tang2020molecular,ohto2019coupling,yu2021bending,du2024aluminumoxide} Understanding these
systems requires molecular simulations that can describe chemical bonding, charge
redistribution, polarization, and finite-temperature sampling within a unified atomistic
framework.

Ab initio molecular dynamics (AIMD) provides such a framework by evaluating electronic
structure along finite-temperature trajectories, and has become indispensable for systems in
which static electronic-structure calculations or empirical force fields are insufficient.~\citep{marx2009aimd,kuhne2014firstprinciples}
However, the cost of solving the electronic structure at every time step restricts direct AIMD
to relatively small systems and short trajectories, limiting its ability to converge ensemble
properties, sample rare events, and describe extended heterogeneous interfaces.~\citep{jia2020pushing} Machine
learning molecular dynamics (MLMD) has emerged as a practical strategy for extending
ab initio accuracy to longer time scales and larger system sizes.~\citep{behler2021fourth,unke2021machine,deringer2021gaussian} In MLMD, machine learning
potentials (MLPs) trained on ab initio reference data replace repeated electronic-structure
calculations during molecular dynamics, preserving a direct connection to first-principles
theory while greatly reducing the cost of sampling.

In this work, we use the term AI-accelerated ab initio (\AItwo{}) methods to refer broadly
to computational workflows in which MLPs, together with related learned property models
when needed, are trained on ab initio data to accelerate molecular simulation, free-energy
sampling, electrochemical modeling, or property prediction. This includes MLMD based on
MLPs and its extensions to free-energy calculations, electrochemical interface simulations,
and AI-assisted spectroscopic prediction. This expanded reach is already changing what
chemists study in practice. In heterogeneous catalysis, for example, \AItwo{}-driven
molecular dynamics is shifting the focus from static reaction energetics to
finite-temperature free energies, revealing the effects of surface restructuring,
adlayer reorganization, defect migration, and active-phase transitions, which are
essential for reliable mechanistic interpretation and quantitative comparison with
experiment.\citep{D4SC05399K,gong2024anie,zhang2024jacs,zhou2024sciadv}

As \AItwo{} methods move from model systems to realistic chemical applications, the main
bottleneck is no longer only the cost of electronic-structure calculations, but also the
complexity of the computational workflow.~\citep{zhang2020dpgen,vandermause2020flare} A practical \AItwo{} study typically requires
iterative data generation, ab initio labeling, model training, configuration exploration,
uncertainty estimation, configuration screening, relabeling, validation, and production
simulation.~\citep{zhang2020dpgen,vandermause2020flare} Advanced applications further multiply this complexity: free-energy calculations
require multiple thermodynamic or alchemical states; electrochemical simulations require
long-range electrostatics, electrode polarization, and often Wannier-center information; and
spectroscopic predictions require learned dipoles, polarizabilities, or related response
properties along long trajectories. Connecting these steps across electronic-structure codes,
machine learning potential packages, molecular dynamics engines, enhanced-sampling tools,
analysis scripts, and high-performance-computing schedulers usually demands substantial
boilerplate code for file conversion, template generation, path management, job submission,
checkpointing, error recovery, and post-processing.

Existing workflow tools address important parts of this challenge, but a gap remains for
complex \AItwo{} research. General-purpose engines such as Parsl, FireWorks, DFlow, and
Snakemake~\citep{babuji2019parsl,jain2015fireworks,dflow2024,molder2021snakemake} provide powerful task orchestration and resource management, but do not encode
the domain-specific operations that recur throughout computational chemistry workflows.
Specialized active-learning tools such as DPGEN and FLARE~\citep{zhang2020dpgen,vandermause2020flare} have greatly advanced machine learning
potential construction, but are primarily organized around specific workflow patterns and
software ecosystems. The increasing diversity of \AItwo{} applications calls for a toolkit that
combines chemistry-aware operations, concise user interfaces, restartable workflow
execution, and extensibility across multiple backends and scientific scenarios.

Here we present ai2-kit, a software toolkit designed to streamline \AItwo{} workflow
development for complex chemical systems. ai2-kit provides high-semantic-density
command-line interfaces and Python APIs for common computational-chemistry operations,
including structure and dataset conversion, batch configuration generation, active-learning
screening, job orchestration, and workflow recovery. Rather than enforcing a single rigid
pipeline, ai2-kit exposes composable building blocks and provides ready-to-use workflow
templates for representative applications, including machine learning potential training,
free-energy perturbation, electrochemical machine learning potentials, and spectroscopic
prediction. These templates are designed to be modified and extended, allowing users to
develop customized \AItwo{} workflows for their own chemical systems, computational
engines, and research objectives.

In the following sections, we first describe the design principles and command-line/Python
interfaces that give ai2-kit its high semantic density, and next introduce TESLA as its central
active-learning workflow for MLP construction. We then demonstrate representative
workflow templates for MLP training, free-energy perturbation, electrochemical machine
learning potentials, and spectroscopy from MLMD, before discussing how ai2-kit supports
workflow portability and extensibility across different software stacks and chemical
applications.

\section{Ai2-kit overview and theoretical background}
\label{sec:program}

\subsection{Design philosophy and overview}
\label{sec:design}

\begin{figure}[!htbp]
  \centering
  \includegraphics[width=\linewidth]{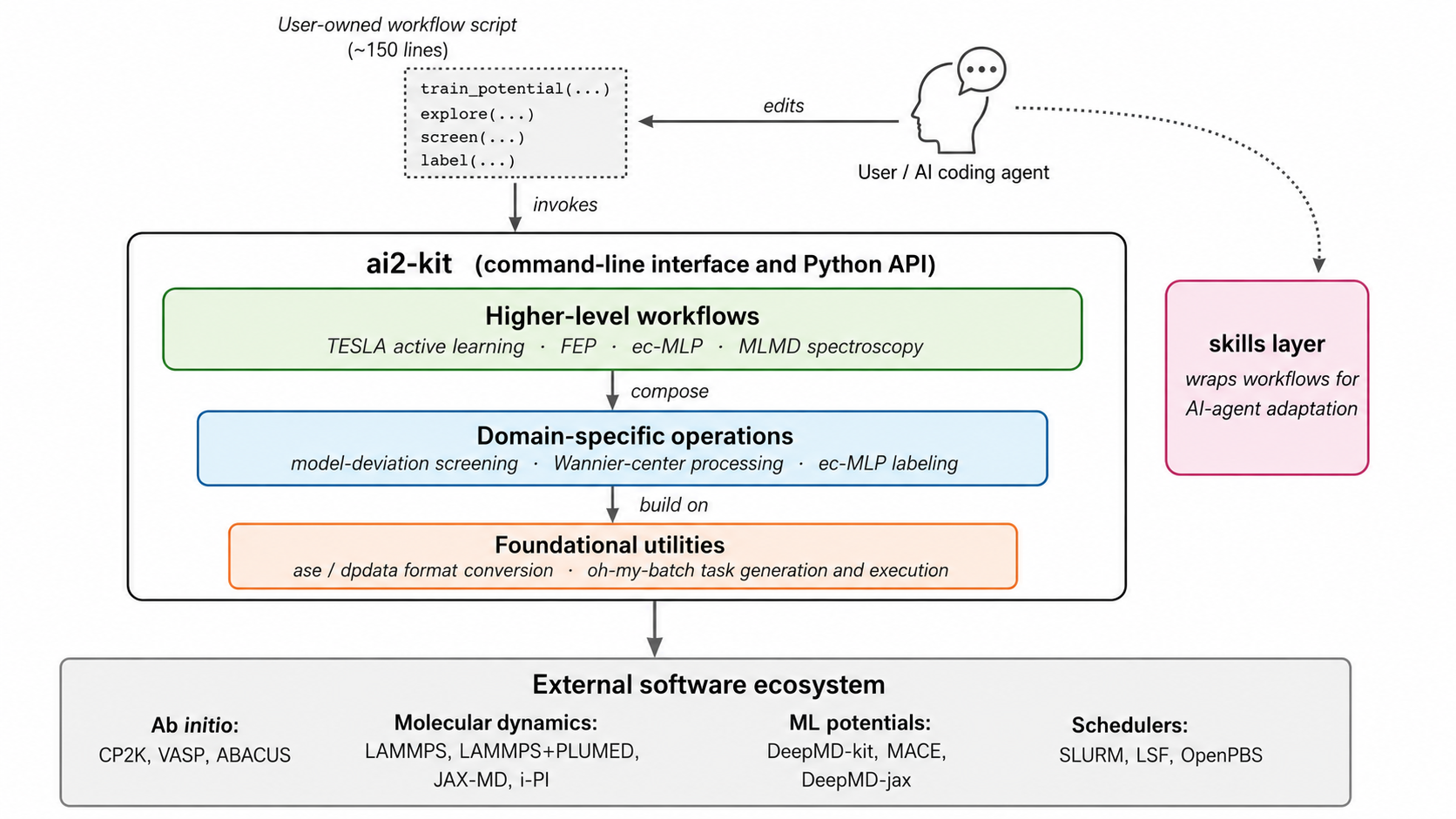}
  \caption{High-level architecture flowchart of ai2-kit.}
  \label{fig:design-philosophy}
\end{figure}

A practical \AItwo{} study for a complex chemical system rarely reduces to a single
command. The researcher must combine an ab initio engine for high-accuracy reference
data, a machine learning framework to train a potential, a molecular dynamics engine for
exploration, often an enhanced-sampling layer to access rare events or alchemical states,
and a chain of supporting scripts that convert formats between stages, generate batches
of jobs, monitor cluster execution, and extract physical observables. Each new system or
methodological choice typically forces fresh decisions at one or more of these points, so
the dominant cost is not raw orchestration but \emph{iteration}: turning a new physical
idea into a running, validated workflow. Existing tools fall short of this. General-purpose
workflow engines such as Parsl, FireWorks, DFlow, and Snakemake~\citep{babuji2019parsl,jain2015fireworks,dflow2024,molder2021snakemake} provide solid task
orchestration but no chemistry-specific operations, so one still needs to write every
format conversion and parameter sweep themselves. Specialized active-learning toolkits
such as DPGEN and FLARE~\citep{zhang2020dpgen,vandermause2020flare} bundle a complete \AItwo{} pipeline behind a configuration interface but
resist customization beyond their prescribed pattern; integrating a new ab initio engine,
modifying the screening criterion, or trying a new sampling scheme usually means editing
internal code.

ai2-kit takes a different stance (Figure~\ref{fig:design-philosophy}). Rather than aiming for full end-to-end automation,
which is impractical given the diversity of \AItwo{} problems encountered in practice, we
lower the cost of iteration by separating engineering plumbing from scientific intent.
The plumbing consists of the recurring infrastructural operations every \AItwo{} workflow
has to perform (data-format conversion, parameter combinatorics, batch job generation,
cluster submission, state recovery) and is provided as a small set of composable
command-line tools with matching Python APIs. The scientific intent (the potential to
train, the sampling strategy to drive, how configurations are screened and labeled, how
the trained model is coupled to downstream simulations) stays in a short, editable
workflow script that the user owns. Adapting a workflow to a new system then
means editing the script and a few input templates, not the underlying tools.

Implemented as a Python package, ai2-kit exposes its functionality through a
command-line interface and a matching Python API, and is internally organized as a
three-tier stack in which each higher tier is a thin composition of the tier below. The
bottom tier provides \emph{foundational utilities}: data- and format-conversion
primitives (built on \texttt{ase} and \texttt{dpdata}) that bridge common ab initio
engines (CP2K, VASP, ABACUS), molecular dynamics engines (LAMMPS, LAMMPS+PLUMED,
JAX-MD, i-PI), and machine learning potential frameworks (DeepMD-kit, MACE,
DeepMD-jax),~\citep{kuhne2020cp2k,kresse1996vasp,zhou2025abacus,plimpton1995lammps,bonomi2009plumed,jaxmd2020,ceriotti2014ipi,wang2018deepmd,Batatia2022mace,Batatia2022Design,deepmdjax2023} together with batch task
generation and execution through \texttt{oh-my-batch} (\texttt{omb}), a command-line
tool distributed with the ai2-kit package that enumerates
parameter combinations, instantiates input and launch templates, and submits jobs to
SLURM, LSF, and OpenPBS schedulers with automatic retry and state recovery. The middle
tier provides \emph{domain-specific operations}: chemistry-aware functionality including
model-deviation- and descriptor-based screening for active learning, Wannier-center
processing for spectroscopic workflows, and electrochemical-interface labeling. The top
tier provides \emph{higher-level workflows} that compose the lower two tiers into the
TESLA active-learning workflow described in the next subsection and its variants for
free-energy perturbation, electrochemical machine learning potentials, and spectroscopy
from machine learning molecular dynamics. A complementary skills layer further allows
AI coding agents to adapt these workflows to new tasks.

\subsection{Program structure}
\label{sec:structure}

ai2-kit realizes the design philosophy above through two interface conventions shared by
all of its tools.
\emph{Multi-files-in-and-out (MFI/O)} lets a single command operate on a whole collection
of files at once: inputs are specified by wildcards and outputs by template variables,
replacing the explicit loops, path bookkeeping, and directory creation that dominate
hand-written workflow scripts.

\emph{Method chaining} lets a sequence of dependent steps compose within a single
command, using a dash (\texttt{-}) as the separator in the command-line interface (CLI)
or chained calls in Python, so a pipeline of reading, transforming, and writing data
proceeds without intermediate variables or files.

\begin{figure}[!htbp]
  \centering
  \includegraphics[width=\linewidth]{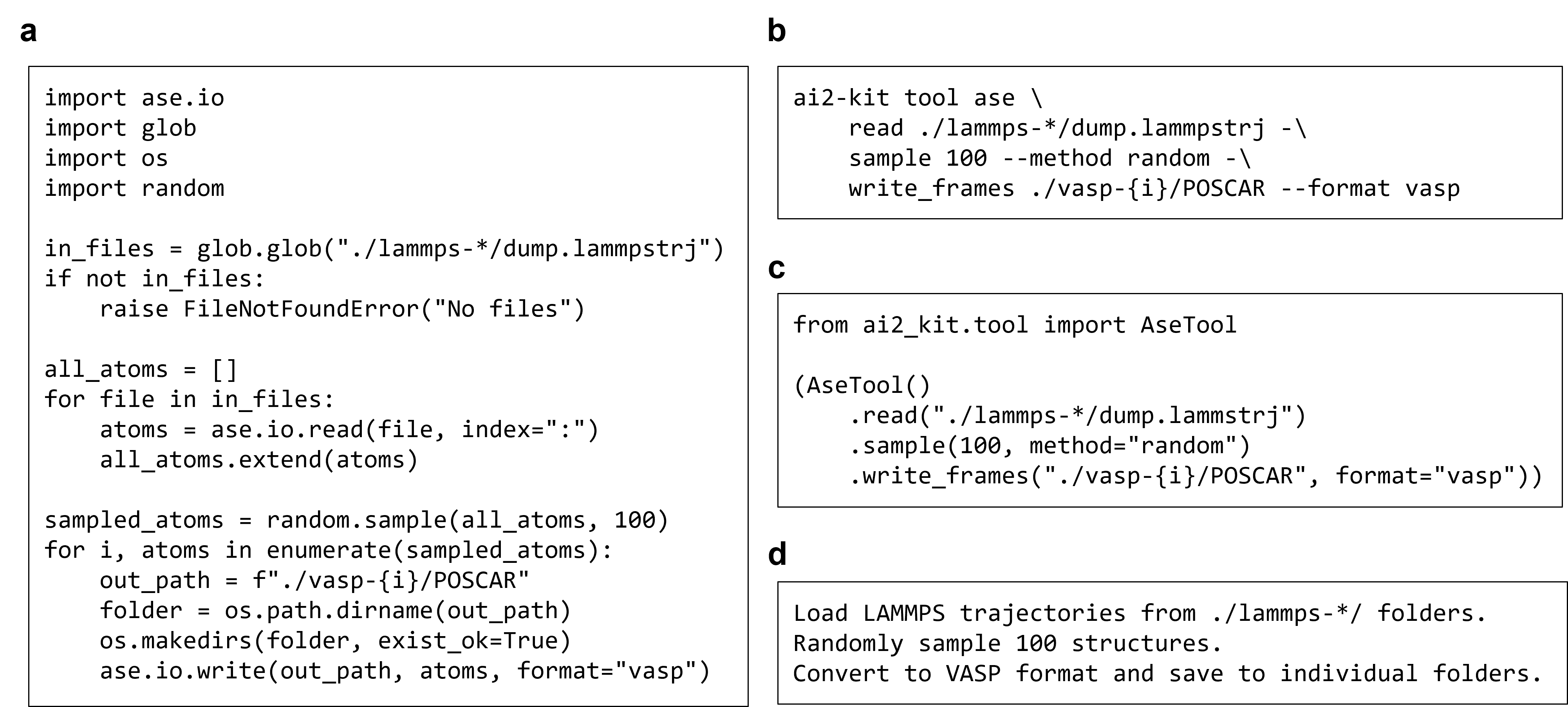}
  \caption{Comparison of four ways to express the same computational task: reading all
    LAMMPS trajectories from \texttt{./lammps-*/}, randomly sampling 100 structures, and
    writing them as VASP POSCAR files in individual task folders. (a) Python
    implementation without ai2-kit (17 lines); (b) ai2-kit command-line interface
    (4 lines); (c) Python implementation with ai2-kit (5 lines); (d) natural-language
    description.}
  \label{fig:semantic-density}
\end{figure}

Together, MFI/O and method chaining give workflow code a natural-language density. A
representative data-handling task that takes 17 lines of explicit Python without
ai2-kit collapses to 4 lines in the ai2-kit CLI or 5 lines in the ai2-kit Python
module, closely matching the line count of a natural-language description
(Figure~\ref{fig:semantic-density}).

The same compression carries over to workflow-level operations. Consider a canonical step
from a TESLA active-learning workflow: enumerating combinations of trained potentials,
initial structures, multiple temperatures and pressures, and unique random seeds, then
generating both LAMMPS input files and submission scripts. The entire step fits in a
single chained command:

\FloatBarrier
\begin{lstlisting}[style=shellstyle]
omb combo \
    add_files     DATA_FILE ./lammps-data/* --abs -\
    add_file_set  DP_MODELS ./deepmd-*/compress.pb --abs -\
    add_var       TEMP      330 430 530 -\
    add_var       PRES      1 2 3 -\
    add_randint   SEED      -n 10000 -a 0 -b 99999 --uniq -\
    set_broadcast SEED -\
    make_files ./lammps-{i}/in.lammps --template in.lammps -\
    make_files ./lammps-{i}/run.sh    --template lmp-run.sh -\
    done
\end{lstlisting}

The hand-written equivalent would span tens of lines of nested loops and template
substitution. The same interface pattern carries through the rest of the toolkit ---
ASE- and dpdata-based wrappers provide broad coverage of common atomic-structure and
DFT-output formats, and \texttt{omb batch} / \texttt{omb job} package the generated tasks
for SLURM, LSF, and OpenPBS schedulers with automatic retry (\texttt{-{}-max\_tries}) and
state recovery (\texttt{-{}-recovery}) on interruption. The full list of principal tools
and their subcommands is documented in the ai2-kit GitHub repository
(\url{https://github.com/chenggroup/ai2-kit}).

A complete TESLA-style workflow, comprising ab initio labeling, ensemble training, biased exploration,
model-deviation screening, and dataset assembly, typically fits in 100--200 lines of
shell script using ai2-kit (see examples in the Github repository). Adapting that skeleton to a new system, potential, or modality requires only localized
edits to the high-level script and a few input templates; the underlying tools are not
touched. Because the workflow scripts are short and self-describing, they also read
naturally to modern coding agents, which we exploit in the section
``\nameref{sec:examples}'' as \emph{skills} to further compress the cost of adapting
workflows to new tasks.

\subsection{TESLA: training--exploration--screening--labeling active-learning workflow}
\label{sec:tesla}

The central workflow provided by ai2-kit is the TESLA
(Training--Exploration--Screening--Labeling Active-learning) workflow, a closed-loop
active learning framework for iteratively constructing training datasets and refining
machine learning potentials. The TESLA workflow is an improved implementation of the DPGEN
approach, redesigned with a more modular architecture and a semantic configuration system
to support diverse \AItwo{} scenarios.

\begin{figure}[!htbp]
  \centering
  \includegraphics[width=0.9\linewidth]{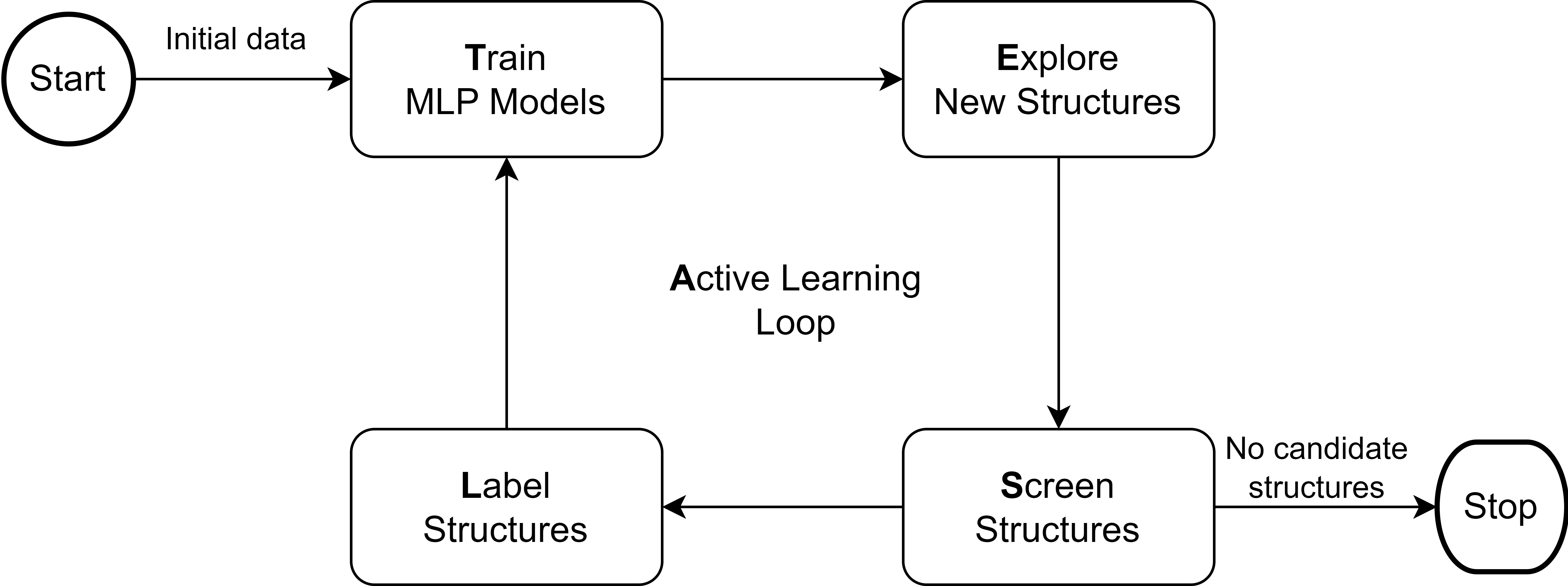}
  \caption{Schematic of the TESLA active-learning workflow. Each iteration cycles
    through four stages --- Training, Exploration, Screening, and Labeling --- forming
    a closed active learning loop.}
  \label{fig:tesla}
\end{figure}

\paragraph{Active learning loop.}
The TESLA workflow operates through iterative cycles, each consisting of four stages
(\Cref{fig:tesla}):
\begin{enumerate}
  \item \textbf{Training.} Multiple MLP models (typically an ensemble of four) are trained
        on the current labeled dataset. The use of multiple models with different random
        initializations provides an estimate of prediction uncertainty in subsequent
        stages. Currently supported training backends include DeepMD-kit, MACE, and
        DeepMD-jax; adopting a different MLP architecture requires only editing the
        training template and the corresponding data-conversion command in the workflow
        script, which we illustrate with a MACE-based example in the section
        ``\nameref{sec:examples}''.
  \item \textbf{Exploration.} The trained models drive molecular dynamics simulations (or
        other structure search methods) to sample the configuration space. During these
        simulations, the disagreement among ensemble members is recorded at each timestep
        as the model deviation. 
  \item \textbf{Screening.} Configurations generated during exploration are classified
        into three categories based on the maximum deviation of atomic forces predicted by
        the ensemble members, denoted \(\sigma_f^{\max}\):
        \begin{itemize}
          \item \emph{Accurate} (\(\sigma_f^{\max} < \sigma_{\text{lo}}\)): the model
                predictions are in good agreement and the configuration is considered
                well described.
          \item \emph{Candidate} (\(\sigma_{\text{lo}} \leq \sigma_f^{\max} <
                \sigma_{\text{hi}}\)): the model shows significant disagreement,
                indicating that this region of configuration space is insufficiently
                represented in the training data. These configurations are selected for
                labeling.
          \item \emph{Failed} (\(\sigma_f^{\max} \geq \sigma_{\text{hi}}\)): the deviation
                is too large, suggesting the configuration lies far outside the domain of
                applicability and could be unphysical or require special treatment.
        \end{itemize}
        The thresholds \(\sigma_{\text{lo}}\) and \(\sigma_{\text{hi}}\) are
        user-specified trust levels. Optionally, candidate configurations can be further
        refined using descriptor-based clustering (SOAP descriptors with DBSCAN) to remove
        structurally redundant configurations and improve the diversity of the selected
        set. Beyond the committee-based model-deviation criterion, ai2-kit also supports
        lightweight uncertainty-quantification methods that operate on a single trained
        model and therefore avoid the cost of training and inferring with a committee, for
        example Last-Layer Prediction Rigidity (LLPR).~\citep{bigi2024rigidity,chong2025rigidities}
  \item \textbf{Labeling.} The selected candidate configurations are computed with ab
        initio methods (CP2K, VASP, ABACUS, etc.) to obtain reference energies and forces.
        These newly labeled data are appended to the training dataset for the next
        iteration.
\end{enumerate}

The workflow continues iterating until a convergence criterion is met, typically
defined as the fraction of \emph{accurate} configurations (the passing rate) exceeding a
predetermined threshold, or until the maximum number of iterations is reached. At each
iteration, the workflow configuration (training hyperparameters, exploration conditions,
etc.) can be automatically updated through a walkthrough table, enabling staged training
strategies such as progressively increasing the number of training steps or expanding the
temperature range of exploration.

\section{Results and discussion}
\label{sec:applications}

The ai2-kit provides ready-to-use workflow examples for a range of \AItwo{} scenarios commonly
encountered in complex chemical studies. In this section, we present four representative
applications --- MLP training, free energy perturbation, electrochemical interface
simulation, and spectroscopy --- demonstrating what ai2-kit can do and how researchers use
it in practice. We then discuss how ai2-kit's design enables easy cross-scenario migration
and an extensible workflow ecosystem.

\subsection{Training a machine learning potential with the TESLA workflow}
\label{sec:mlff}

In modern atomistic molecular simulation, an accurate machine learning potential (MLP)
is a prerequisite for studying a wide range of systems --- from bulk liquids and
crystalline solids to complex interfaces and disordered materials such as amorphous
carbon. Because it is generally not known \emph{a priori} which configurations are needed
to sample the relevant regions of the potential energy surface, constructing such a
potential can be efficiently done with the TESLA active-learning workflow described in
the previous section. ai2-kit provides two ready-to-use entry points into
this workflow ---
a self-contained shell-script example and a configuration-driven Python CLI --- and
supports commonly used MLP frameworks including DeepMD-kit, MACE, and DeepMD-jax.

The shell-script example lives at \texttt{example/use-case/tesla/} in the ai2-kit GitHub
repository and is implemented in
approximately 130 lines of shell built around \texttt{ai2-kit} and the companion tool
\texttt{oh-my-batch}. For a new system, the user can just place an AIMD trajectory at
\texttt{00-config/aimd.xyz}, edit the
four template files under \texttt{00-config/} to specify the system's \texttt{type\_map},
atomic masses, and CP2K basis sets, and updates the three \texttt{slurm-header.sh} files
to match the target cluster's partitions and module environment. The full active-learning
loop, three iterations each cycling through DeepMD training of four models, LAMMPS
exploration at multiple temperatures, model-deviation screening, and CP2K labeling, is
then launched with a single command:

\begin{lstlisting}[style=shellstyle]
./run.sh
\end{lstlisting}

Each stage writes a \texttt{.done} sentinel as it completes, so an interrupted run simply
resumes on re-invocation.

Users who prefer a fully declarative setup can instead invoke the same active-learning
loop through the Python CLI. The 64-water example documented in
\texttt{doc/manual/cll-workflow.md} of the ai2-kit GitHub repository illustrates the pattern, in which the user specifies
data paths, SSH/Slurm settings, model hyperparameters, and model-deviation thresholds step
by step through a small set of YAML files; further details and the launch command are
provided in the ai2-kit repository.

We demonstrate the workflow on two chemically distinct systems: bulk water as a canonical
benchmark, and an Au--CO$_2$ dynamic catalytic system as a more demanding case.

\begin{figure}[!htbp]
  \centering
  \includegraphics[width=\linewidth]{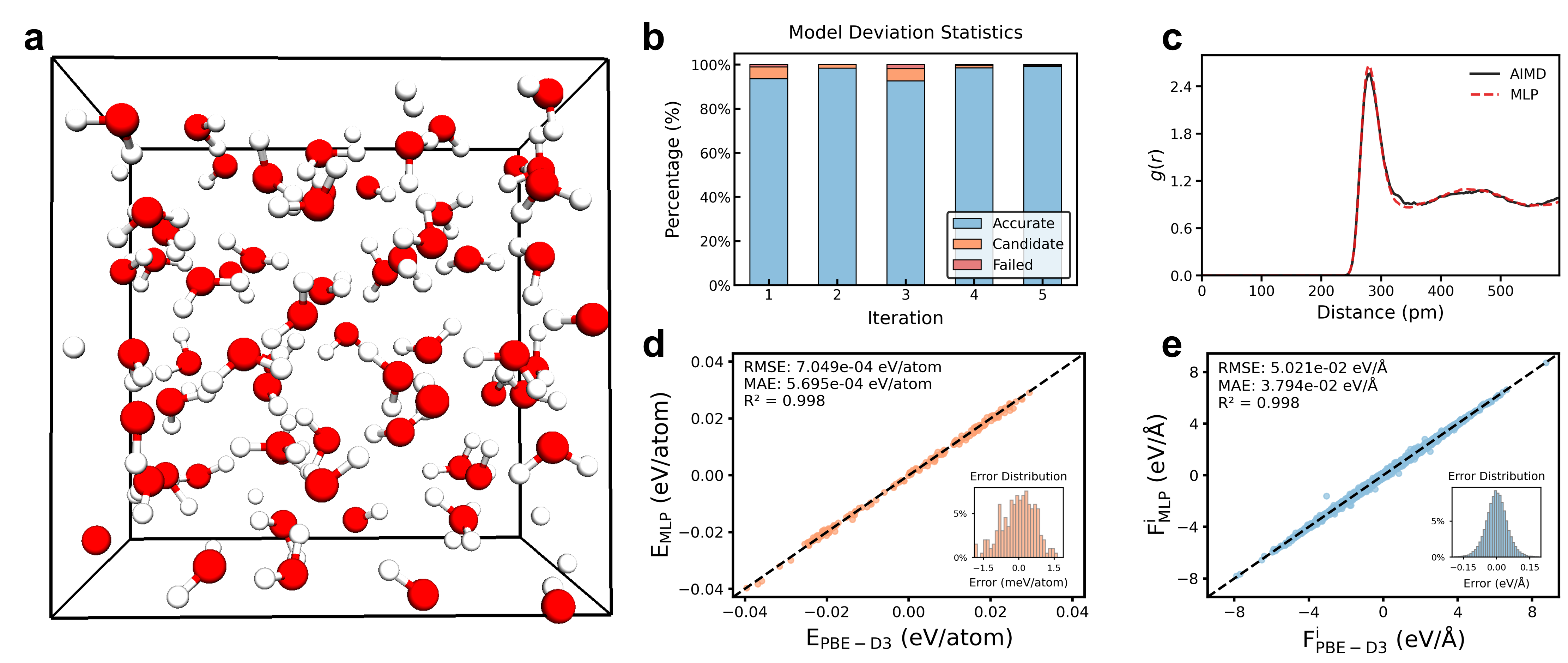}
  \caption{(a) Representative snapshot of the bulk water system. (b) Evolution of the
    model-deviation statistics during the TESLA workflow. (c) Radial distribution function
    (O--O) computed from MLMD (red dashed line) overlaid with the AIMD reference (black
    solid line). (d,e) Parity plots comparing the MLP-predicted (d) energies and (e)
    atomic forces against the DFT reference. The insets display their respective error
    distributions.}
  \label{fig:water-tesla}
\end{figure}

\paragraph{Bulk water.}
As a widely adopted benchmark system in atomistic simulations, bulk water provides an
ideal testbed to demonstrate the efficiency of \texttt{ai2-kit}. Using the TESLA workflow,
we trained an MLP for this fundamental system (\Cref{fig:water-tesla}a). Driven by the
workflow's active-learning strategy, training converged rapidly within just five
iterations, requiring a remarkably small dataset of only $\sim$170 configurations. As
shown in \Cref{fig:water-tesla}b, the proportion of ``accurate'' configurations quickly
stabilized near 100\%, indicating efficient configurational-space exploration. The
resulting MLP exhibits excellent structural fidelity, with the MLMD-derived O--O radial
distribution function in close agreement with the AIMD reference
(\Cref{fig:water-tesla}c). The close agreement between MLP and DFT values on the training set 
indicates that the model has successfully learned the reference data. (\Cref{fig:water-tesla}d, e). These results
demonstrate that \texttt{ai2-kit} can fully automate the generation of reliable and
transferable potentials for a canonical liquid with minimal data and computational cost.

\paragraph{Au--CO$_2$ catalysis.}
We next apply the workflow to a more chemically
demanding system, catalysis on metal nanoclusters. As discussed in the introduction, the
chemistry of such systems often depends not on static reaction energetics alone but on
finite-temperature free energies that capture phenomena such as cluster restructuring
and phase transitions of the active site. The Au--CO$_2$ system is a
representative model for studying gold-based
catalysis, focusing on the interactions of CO$_2$, CO, and oxygen with gold clusters. The
relevant elementary processes include CO$_2$ adsorption and dissociation on the cluster
surface, followed by surface diffusion of oxygen atoms and adsorbed CO molecules. These
processes are closely related to CO oxidation and are central to understanding the
activity and selectivity of gold catalysts. For sub-nanometer gold clusters, the cluster
structures are strongly dynamic at finite temperature; adsorption and association of CO
and oxygen can further alter the configurational evolution and phase-transition behavior
of the clusters.\citep{sun2019phase}

Previous studies have shown that, within certain temperature ranges, different reaction
states may correspond to distinct cluster structures and melting behavior, thereby
changing the effective melting temperature of the system.\citep{sun2019phase,fan2024dynamic} Because the solid-like and
liquid-like phases have substantially different entropies, such phase transitions can
strongly affect the free energy of CO$_2$ dissociation and produce an apparent
free-energy jump near specific temperatures. The Au--CO$_2$ system therefore captures
fundamental reaction processes in gold-based catalysis while also providing an ideal
model for investigating dynamic catalytic effects in small metal clusters. 

\begin{figure}[!htbp]
  \centering
  \includegraphics[width=\linewidth]{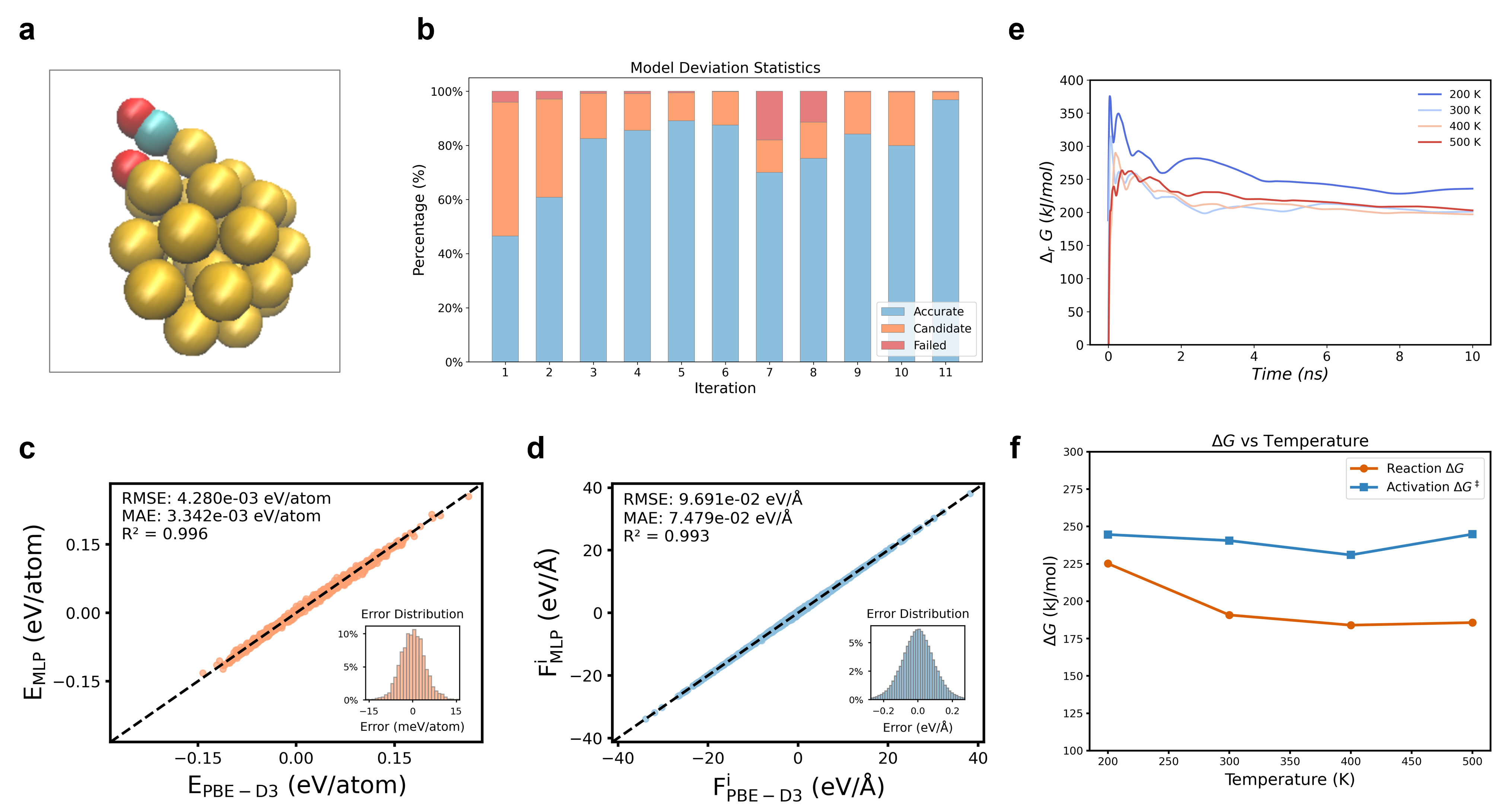}
  \caption{(a) Representative snapshot of the Au/CO$_2$ system. (b) Evolution of the
    model-deviation statistics during the TESLA workflow. (c,d) Validation accuracy of the
    trained MLP model, showing (c) energy and (d) force predictions against reference data.
    (e) Convergence of the free-energy difference at 200, 300, 400, and 500~K.
    (f) Temperature dependence of the free energy.}
  \label{fig:auco-tesla}
\end{figure}

Building on TESLA, ai2-kit provides a comprehensive example for this Au--CO$_2$ catalytic
system, specifically targeting the dissociation of a single CO$_2$ molecule --- the
reverse of CO oxidation --- catalyzed by a 38-atom Au cluster. The example includes TESLA
workflow scripts and configuration templates for structural sampling and potential
training, analysis scripts for evaluating the fitting accuracy of the trained potential,
and post-processing scripts for computing both reaction and activation free energies
(\Cref{fig:auco-tesla}). During the active-learning loop, the first five iterations
use shorter exploration trajectories and smaller labeling batches, with 30,000 MD steps
and 100 newly labeled configurations per iteration. Starting from the sixth iteration, the
workflow switches to longer exploration and larger labeling batches, with 100,000 MD steps
and 200 labeled configurations per iteration. The model-deviation statistics show that
the ensemble of concurrently trained models becomes progressively more consistent during
the first five iterations. After the sixth iteration, the model disagreement increases
because the exploration and labeling scale is expanded; the effect of the increased
labeling set appears from the seventh iteration onward, after which the models become
increasingly consistent again, indicating that the configurational sampling has become
sufficiently comprehensive.

Beyond the two illustrative examples above, ai2-kit has also been adopted in recent
\AItwo{} studies spanning electrochemistry, catalysis, and materials discovery. In
electrochemistry, ElectroFace used ai2-kit to help organize and standardize an
AI-accelerated \emph{ab initio} molecular dynamics dataset for electrochemical
interfaces, improving the accessibility and reuse of interfacial simulation
data.\citep{zhuang_artificial_2025} In dynamic catalysis, the CatFlow workflow built on
ai2-kit to automate machine learning potential training and free-energy calculations for
catalytic reactions, thereby reducing the manual effort required for end-to-end
finite-temperature reaction profiling.\citep{liu_catflow_2025} In semiconductor alloy
discovery, ChecMatE combined ai2-kit with stochastic surface walking (SSW)
methods\citep{shang_stochastic_2013,shang_stochastic_2014} to automatically generate
machine learning potentials and phase diagrams, showing that ai2-kit can also support
global structural exploration beyond conventional molecular-dynamics-based
sampling.\citep{guo_checmate_2023}

\subsection{Free energy perturbation for reversible particle insertion}
\label{sec:fep}

Computing thermodynamic properties such as redox potentials, pKa values, and solvation
free energies from molecular simulations requires free energy perturbation (FEP) methods
that sample multiple thermodynamic states connected by a coupling parameter \(\lambda\).
Training machine learning potentials (MLPs) for FEP calculations presents a distinct
challenge: the potential must accurately describe not only the physical endpoints but also
the intermediate states along the alchemical transformation pathway.

\begin{figure}[!htbp]
  \centering
  \includegraphics[width=\linewidth]{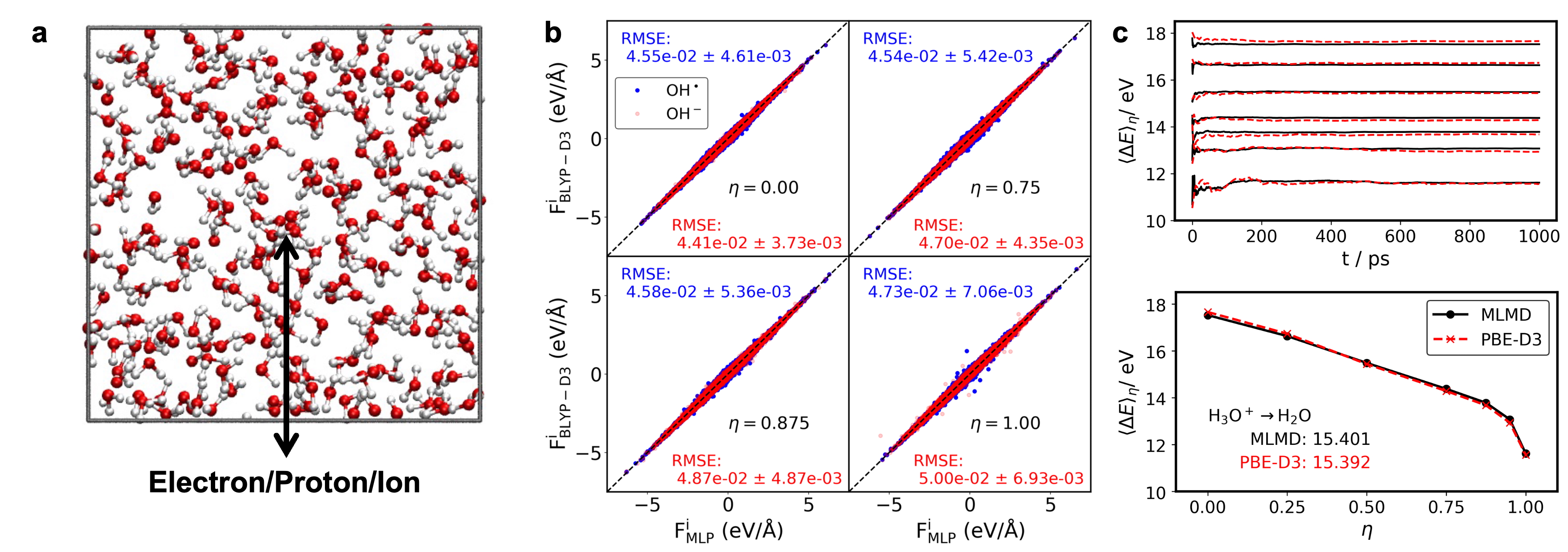}
  \caption{(a) Schematic of the generalized particle insertion framework for electron,
    proton, and ion transfer processes. (b) Validation of redox free energy calculations
    using the OH/OH\textsuperscript{$-$} redox couple. Force errors for both initial
    and final states remain strictly bounded across different coupling parameters.
    (c) Validation of deprotonation free energy for the hydronium ion
    (H\textsubscript{3}O\textsuperscript{+}). Additional dense sampling near the endpoint
    (\(\eta = 0.90, 0.95\)) ensures high accuracy, yielding an error of less than
    0.05~eV compared to DFT validation.}
  \label{fig:fep}
\end{figure}

To address this, ai2-kit supports FEP calculations by extending the TESLA workflow to
seamlessly handle the dual-state nature of free energy surfaces. As illustrated in the
generalized particle (electron/proton/ion) insertion framework
(\Cref{fig:fep}a), ai2-kit introduces free energy perturbation directly into the
active learning loop. Unlike a standard TESLA loop driven by a single data stream, the FEP
workflow constructs coupled potential energy surfaces to systematically collect datasets
for both the initial and final states. Users can implement this by configuring separate
input templates for the two states and utilizing ai2-kit's path partitioning to isolate
the data streams, requiring minimal modifications to the standard setup.

The workflow handles different types of chemical transformations through distinct
potential mapping strategies (\Cref{fig:fep}a). For electron transfer processes, such
as redox free energy calculations, the same atomic structure must be computed in both
oxidized and reduced electronic states. These states require different \emph{ab initio}
configurations due to differing charges and multiplicities. ai2-kit manages this by
accepting separate labeling templates for each state and training two independent MLPs.
During the exploration stage, LAMMPS hybrid potentials linearly mix the two state
potentials at a series of \(\eta\) values.

Conversely, for proton or ion transfer processes like pKa and solvation calculations, the
protonated and deprotonated structures differ explicitly in atomic coordinates. Therefore,
a single universal potential set suffices for both states. During MD exploration,
DeePMD-kit exclusion parameters are dynamically applied to selectively deactivate
interactions involving the transferred particle (e.g., the transferring proton),
effectively yielding the deprotonated-state potential. At the labeling step, ai2-kit's
format conversion tool automatically extracts and generates the corresponding
configurations from the MD output structures. Under this framework, ai2-kit fully
automates the complexity of dual data streams, multi-\(\eta\) sampling, state-specific
labeling, and format conversions --- operations that are traditionally highly manual and
error-prone.

To demonstrate the validity of this training methodology, we evaluated the workflow on
both electron and proton transfer processes. For redox free energy, we investigated the
OH/OH\textsuperscript{$-$} redox couple (\Cref{fig:fep}b). By employing the
dual-potential approach, the workflow achieves high precision across the entire alchemical
pathway. The force and energy errors remain strictly bounded at different coupling
parameters (\(\eta\)), confirming that the coupled potential energy surfaces accurately
capture the electronic transition between the radical and the anion.

For proton transfer reactions, we computed the deprotonation free energy of the hydronium
ion (H\textsubscript{3}O\textsuperscript{+}) (\Cref{fig:fep}c). The workflow
successfully sampled the transformation using a single MLP with interaction exclusions.
While the standard \(\eta\) windows (0.00, 0.25, 0.50, 0.75, 1.00) provided the baseline
sampling, ai2-kit's combinatorial tools easily facilitated the addition of dense sampling
at \(\eta = 0.90\) and \(0.95\) to properly integrate the sharp gradient changes near the
endpoint. The resulting free energy profile demonstrates excellent quantitative agreement
with \emph{ab initio} benchmarks, yielding an error of less than 0.05~eV between the
MLMD-derived deprotonation free energy and direct DFT validations.

Wang \emph{et al.} previously applied an earlier iteration of this workflow to compute
redox potentials and pKa values for a series of organic molecules in aqueous organic redox
flow batteries,~\citep{wang2022automated,wang2024redox} using DeePMD-kit potentials with LAMMPS hybrid-potential MD and CP2K
labeling.

\subsection{Incorporating electric field in simulating electrochemical interfaces}
\label{sec:ecmlp}

Simulating electrochemical interfaces, such as electrode--electrolyte systems in
batteries, electrocatalysis, or corrosion, requires accurate treatment of long-range
electrostatic interactions and dielectric response, which standard machine learning
potentials struggle to capture. To address this, ai2-kit provides a variant of the TESLA
workflow that trains an electrochemical machine learning potential (ec-MLP). The ec-MLP
framework combines a machine-learning-based Wannier center model, which predicts the
polarization response of the electrolyte via atomic dipole moments, with a polarizable
electrode treatment that resolves the induced charge distribution on the electrode under
applied bias. Together, these components enable efficient and accurate description of
long-range electrostatics and dielectric response at electrified interfaces, making ec-MLP
well suited for electrochemical simulations of battery interphases, electrocatalytic
surfaces, and corrosion processes.

The ec-MLP variant of TESLA augments the base active-learning loop with an additional
Wannier-center training branch. At the labeling stage, the input templates for a
first-principles engine such as CP2K, VASP, or Mokit\citep{mokit2026} must be explicitly configured to
sample Wannier centers alongside energies and forces; ai2-kit then converts these outputs
--- including center coordinates, spreads, and charge assignments --- into the format
required by the DeepMD-kit Deep Wannier module~\citep{zhang2020deepwannier} with a single command. Furthermore, the
training configuration necessitates a sequential two-step setup: a Deep Wannier model is
trained prior to the ec-MLP, and its predicted atomic dipoles are passed as input features
to the ec-MLP training step. For the exploration stage, ai2-kit couples the trained models
with LAMMPS running in constant-potential mode~\citep{ahrens2022electrode}: the ec-MLP provides short-range energies
and forces, Deep Wannier predicts Wannier center positions on the fly, and a polarizable
electrode method self-consistently determines electrode atomic charges that satisfy the
applied bias. Despite these additional configuration steps, the complete ec-MLP workflow
is launched with a single command in the same style as the standard TESLA workflow; the
corresponding template files and launch script are provided in the ai2-kit repository.

\begin{lstlisting}[style=shellstyle]
ai2-kit tool dpdata read ./iter-00*/cp2k/job-*/ --fmt cp2k/dplr \
    --cp2k_output="output" --wannier_file="wannier.xyz" \
    --type_map="[Cu,H,O]" --sel_type="[2]" --model_charge_map="[-8]" \
    - write ./new-dataset
\end{lstlisting}

\begin{figure}[!htbp]
  \centering
  \includegraphics[width=0.9\linewidth]{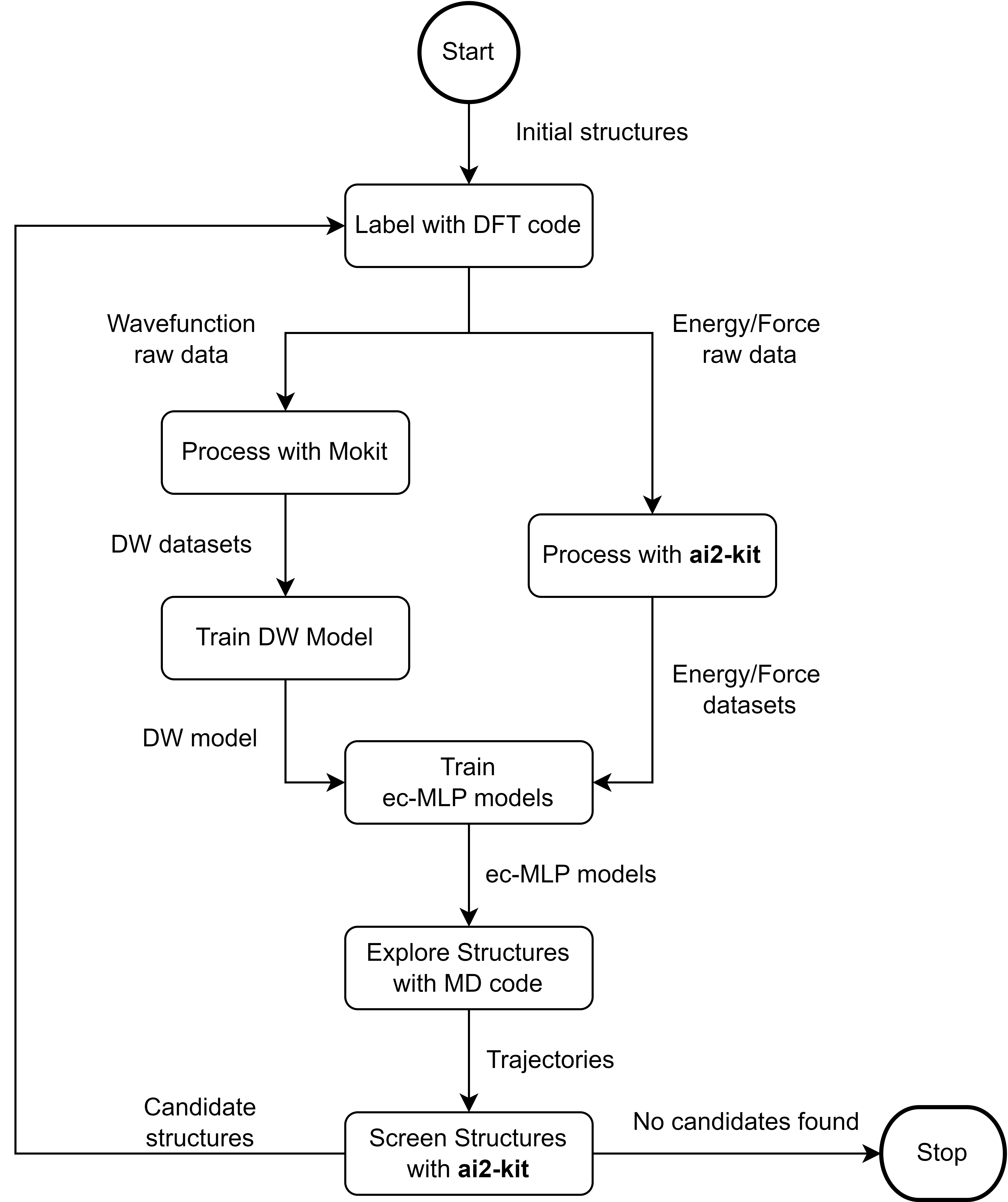}
  \caption{Schematic of the ec-MLP active-learning workflow in ai2-kit. The loop
    partitions the learning process into two coupled branches: a Deep Wannier (DW) model
    is first trained to predict atomic dipoles for long-range interactions, and its
    outputs are embedded as input features to train the ec-MLP model for short-range
    interactions. Together, they drive constant-potential MD exploration to sample and
    screen new candidate structures.}
  \label{fig:ecmlp-workflow}
\end{figure}

\begin{figure}[!htbp]
  \centering
  \includegraphics[width=\linewidth]{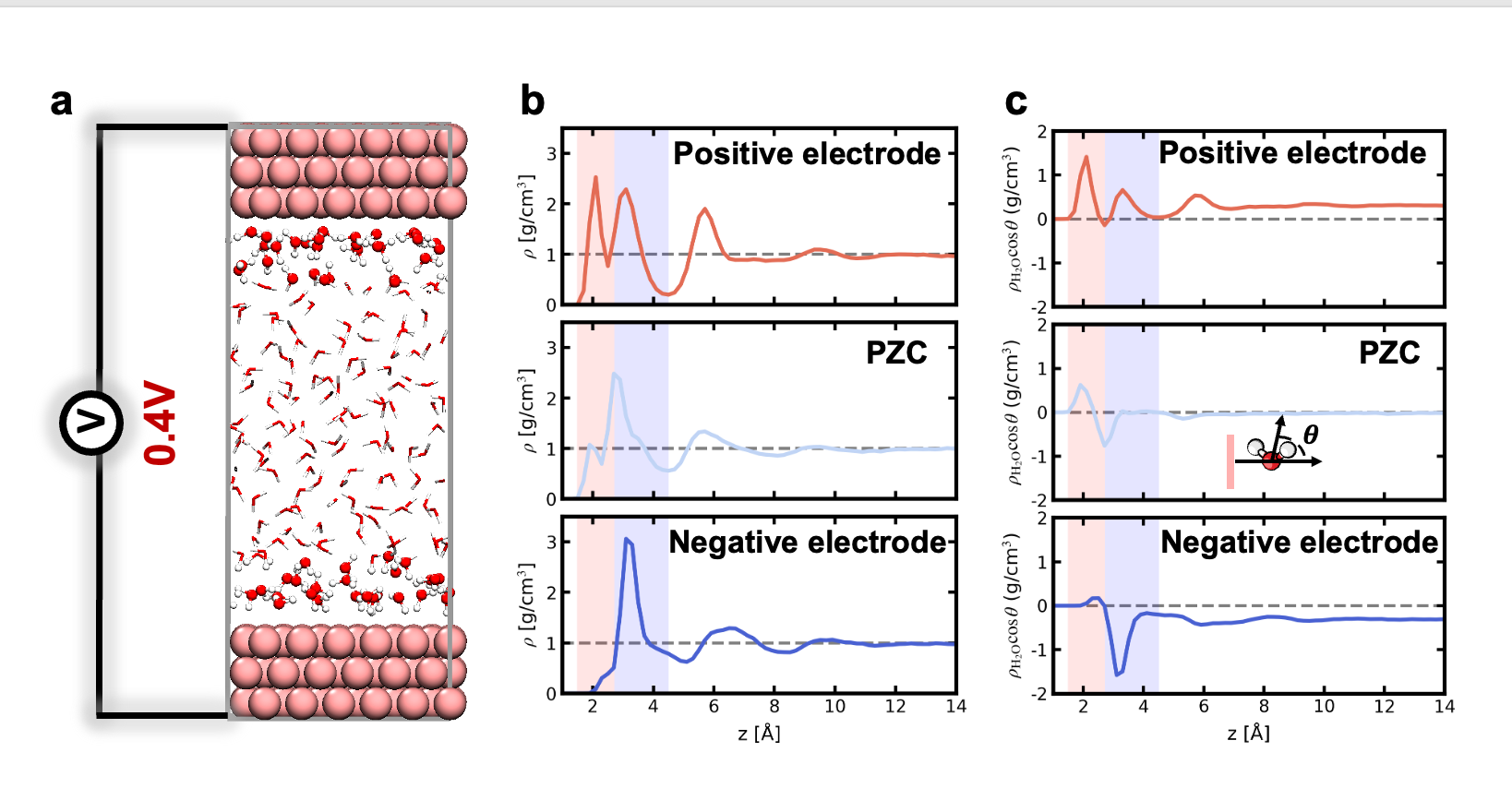}
  \caption{Demonstration of the ec-MLP workflow applied to the Cu(111)/water interface
    under constant-potential MD simulations. (a) Schematic representation of the
    simulation cell subjected to an external bias of 0.4~V. (b) Water density profiles
    (\(\rho\)) near the positive electrode, at the potential of zero charge (PZC), and
    near the negative electrode. The red and blue shaded areas highlight the chemisorbed
    and physisorbed water layers, respectively. (c) Corresponding dipole density
    profiles (\(\rho_{\text{H}_2\text{O}} \cos \theta\)).}
  \label{fig:ecmlp}
\end{figure}

As a demonstration of this capability, we applied the ec-MLP workflow to the Cu(111)/water
interface under constant-potential MD simulations. \Cref{fig:ecmlp}a illustrates the
simulation cell subjected to an external bias of 0.4~V. The high precision of the ec-MLP
ensures stable MD trajectories that reliably capture voltage-induced structural
transitions at the electrochemical interface. \Cref{fig:ecmlp}b displays the water density
profiles (\(\rho\)) near the positive electrode, at the potential of zero charge (PZC),
and near the negative electrode. The shaded red and blue regions clearly demarcate the
distinct formation of chemisorbed and physisorbed water layers, respectively. This
interfacial restructuring is accompanied by pronounced dipolar reorientation, as depicted
by the water orientation profiles (\(\rho_{\text{H}_2\text{O}} \cos \theta\)) in
\Cref{fig:ecmlp}c. The ec-MLP faithfully captures the specific orientation of water
molecules within the chemisorbed and physisorbed regions in response to the local
electrode polarization. Additionally, the orientation bias extending into the bulk region
reflects the absence of ionic screening in pure water, where the unscreened electric field
drives long-range dipole alignment.

Full methodological details and an application of this workflow to the more complex
Pt(111)/aqueous NaCl interface were reported by Zhu and Cheng;~\citep{ec-MLP}
their simulated differential capacitance agreed quantitatively with experimental measurements.

\subsection{Validating complex structures by computing spectroscopies}
\label{sec:spectra}

Spectroscopy provides a natural bridge between atomistic simulations and experiments:
computed IR, Raman, and sum-frequency generation (SFG) spectra can be directly compared
with laboratory measurements, offering both a stringent test of the underlying model and a
way to interpret experimental features in molecular terms. Obtaining converged spectra,
however, requires long trajectories and large simulation cells to adequately sample the
relevant conformational and dielectric fluctuations. In this regime, direct ab initio
molecular dynamics becomes prohibitively expensive and is usually limited to short
trajectories that yield noisy, poorly converged spectra. Machine learning molecular
dynamics (MLMD) closes this gap --- with a machine-learned potential (MLP) trained to ab
initio accuracy, nanosecond-long trajectories of thousands of atoms become routine,
enabling reliable spectroscopic predictions that match the resolution of experimental
data.

ai2-kit provides a dedicated spectroscopy workflow for calculating IR, Raman, and SFG
spectra, and the same machinery can be applied to other properties derived from dipole
moments and polarizabilities. The workflow is organized in two stages. A TESLA-trained MLP
first drives large-scale MLMD sampling of the conformational ensemble at ab initio
accuracy. Then, Deep Wannier (DW) models~\citep{zhang2020deepwannier,sommers2020raman},
trained on Wannier center outputs from the sampled configurations --- computed both at
zero field and with small electric fields applied along \(x\), \(y\), and \(z\) ---
predict atomic dipoles and polarizabilities along the full MLMD trajectory. Ensemble
averages, time-correlation functions, and their Fourier transforms yield the final
spectra. ai2-kit handles sampling, format conversion for spectroscopic codes, batch
Wannier/polarizability calculations, and ensemble-averaged property computation as a set
of composable commands.

\begin{figure}[!htbp]
  \centering
  \includegraphics[width=0.9\linewidth]{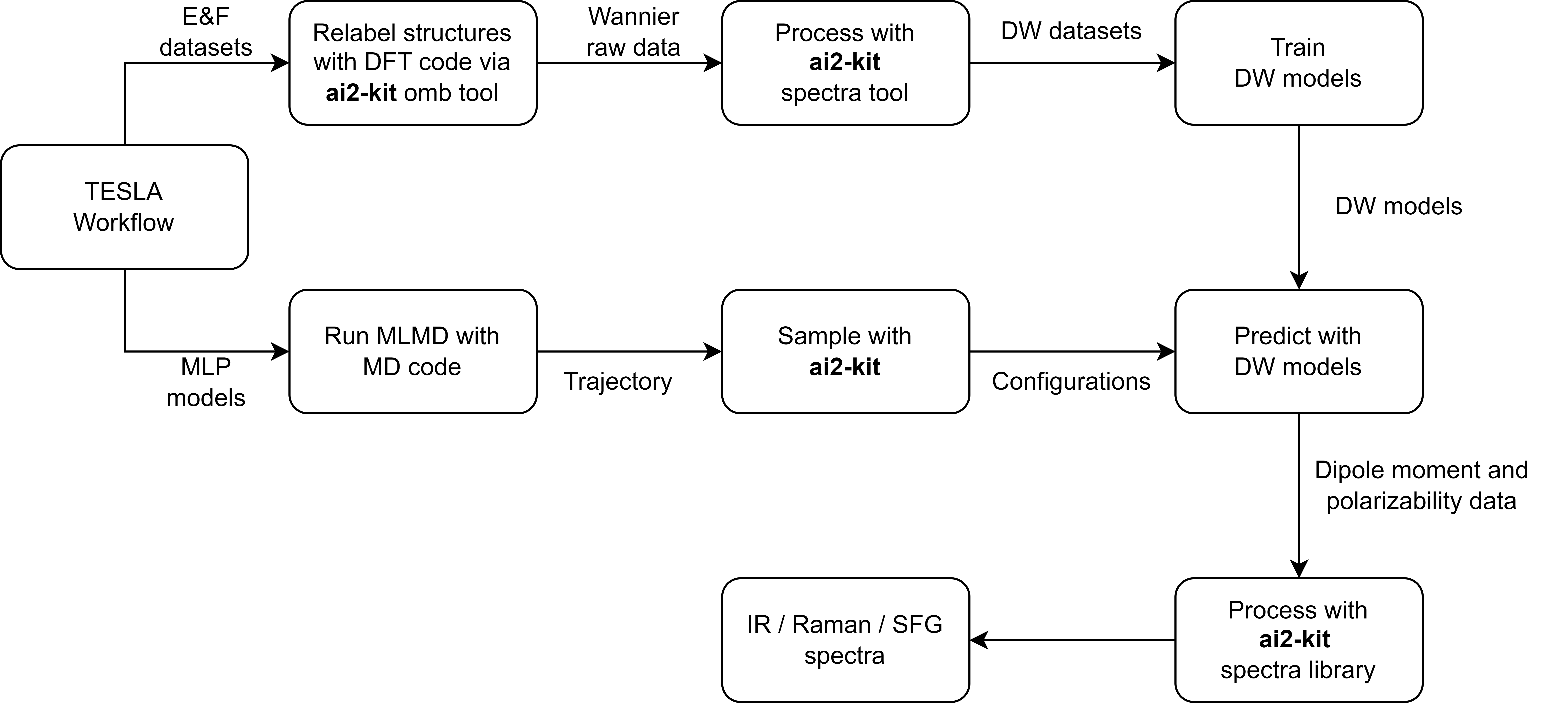}
  \caption{Schematic of the ai2-kit spectroscopy workflow. A TESLA-trained MLP drives
    large-scale MLMD sampling; a Deep Wannier (DW) model trained on Wannier-center
    outputs (at zero field and with small applied fields along \(x\), \(y\), \(z\))
    predicts atomic dipoles and polarizabilities along the full trajectory;
    ensemble-averaged time-correlation functions and their Fourier transforms yield the
    final IR/Raman/SFG spectra.}
  \label{fig:sfg-workflow}
\end{figure}

To illustrate this capability, we demonstrate the calculation of the IR, Raman and SFG spectra at the
air--water interface (\Cref{fig:sfg-workflow}). The corresponding shell-script example is
available in the ai2-kit repository. This script builds on ai2-kit's CLI tools and
consists of the following main steps. First, the TESLA loop is used to train an MLP for
the air--water system. Next, ai2-kit's \texttt{omb} command automates Wannier-center
calculations on the sampled configurations: one at zero field and three with small
electric fields applied along the \(x\), \(y\), and \(z\) directions --- the finite-field
calculations needed to extract polarizabilities via finite differences. Finally, the
script calls \texttt{dp} to train two DW models, one for dipole moments and one for
polarizabilities.

Once the trained MLP and DW models are available, users can perform the production MLMD
and compute their target spectra from the resulting trajectories and per-frame dipole and
polarizability outputs. Because this stage is post-processing rather than part of
ai2-kit's core workflow management, we do not detail it here; reference scripts for IR,
Raman, and SFG analysis are nonetheless provided in the ai2-kit repository.

\begin{figure}[!htbp]
  \centering
  \includegraphics[width=\linewidth]{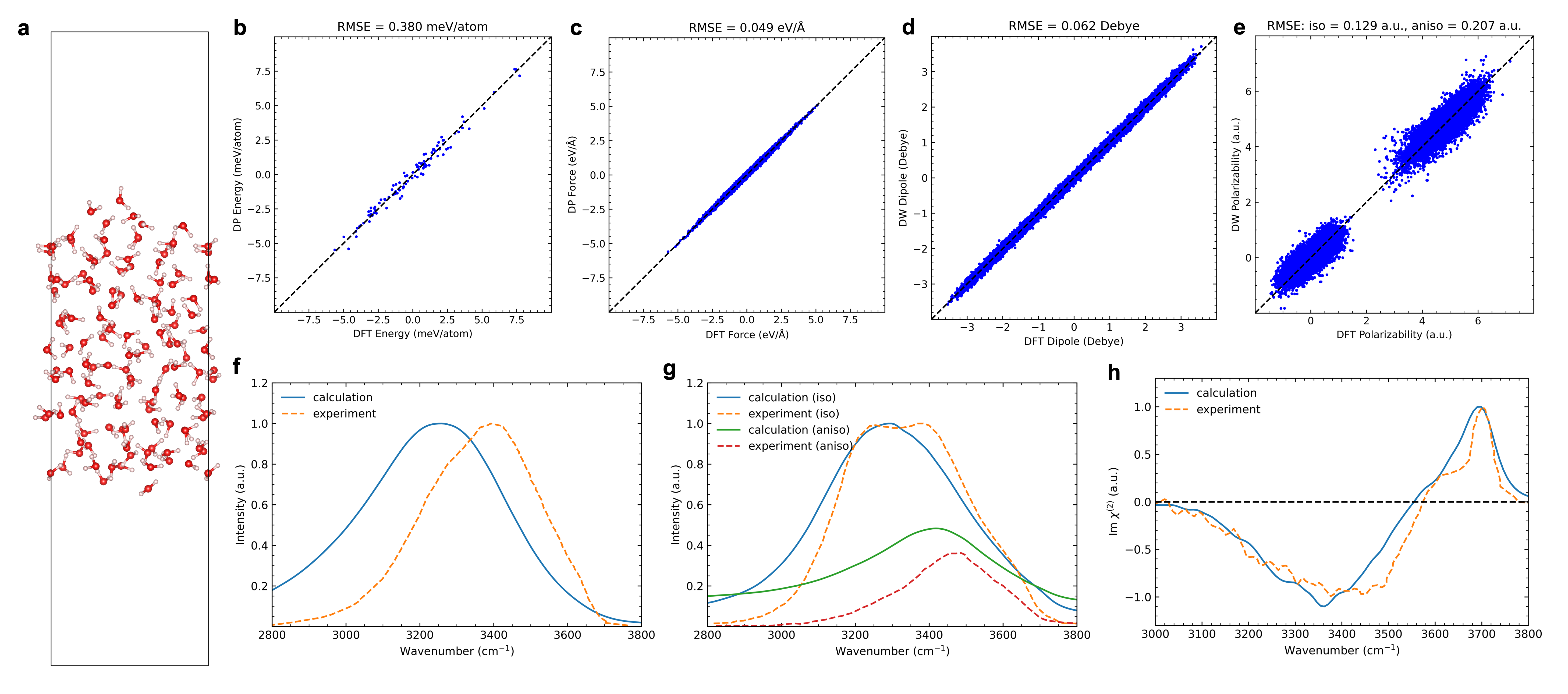}
  \caption{Validation of the trained DP and DW models for the air--water interface.
  (a) Snapshot of the simulated air--water interface with a vacuum region representing the air phase.
  (b) Parity plot of the predicted energy per atom against the DFT reference values. The average value of the
  energy of was shifted to zero for better visualization.
  (c) Parity plot of the predicted atomic forces against the DFT reference values.
  (d) Parity plot comparing the DW-predicted dipole moments with the DFT reference values. 
  (e) Parity plot comparing the DW-predicted polarizability tensor components with the DFT reference values.
  (f) Calculated and experimental~\citep{larouche2008ir} IR spectra for the bulk-like region. 
  (g) Calculated and experimental~\citep{scherer1974raman} isotropic and anisotropic Raman spectra for the bulk-like region.
  (h) Calculated and experimental~\citep{wang2024sfg} imaginary SFG spectra for the interfacial region,
  including the xxz and yyz polarization components.}
  \label{fig:fig10_spectra_results}
  \end{figure}

Results for the air--water interface are shown in \Cref{fig:fig10_spectra_results}. A structural
snapshot of the simulated slab, with a vacuum region representing the air phase, is
depicted in \Cref{fig:fig10_spectra_results}a. To validate the accuracy
of the trained DP and DW models, 100 configurations were randomly selected from a 
2 ns trajectory, and the energies, forces, dipole moments, and polarizabilities 
were recalculated using reference DFT calculations for comparison. The results are shown in 
\Cref{fig:fig10_spectra_results}b--e, respectively. The RMSEs of the predicted energy per atom and atomic forces are 
below 1 meV/atom and 0.1 eV/\AA, respectively, satisfying the general accuracy requirements for 
machine-learning potentials. The predicted dipole moments are in good agreement with the DFT reference values, 
whereas the polarizability tensor components show a larger scatter. Nevertheless, 
as demonstrated below, these deviations do not significantly affect the calculated vibrational spectra.

We then calculated the IR spectrum and the isotropic and anisotropic Raman spectra for the 
bulk-like region of the air--water interface, together with the imaginary parts of the xxz and 
yyz polarization components of the SFG spectrum for the interfacial region. For clarity, 
only the O--H stretching region is compared with the corresponding experimental spectra, 
as shown in \Cref{fig:fig10_spectra_results}f--h. The calculated Raman and SFG spectra reproduce both the peak 
shapes and vibrational frequencies observed experimentally, demonstrating that the present 
workflow can efficiently obtain vibrational spectra with DFT-level accuracy. The main discrepancy 
appears in the IR spectrum. This difference arises because a uniform nuclear quantum correction 
factor of 0.96 ~\citep{tang2020molecular} was applied to the calculated vibrational frequencies to partially account for 
the missing nuclear quantum effects in the classical simulations of O--H stretching vibrations. 
This correction brings the Raman and SFG peak positions into better agreement with experiment, 
while causing a slight deviation in the IR peak frequency. Since the purpose of this validation 
is to demonstrate the feasibility and accuracy of the computational workflow rather than to achieve 
exact quantitative agreement with all experimental spectra, this minor discrepancy does not affect the overall conclusion.

Beyond the air--water interface, the same workflow is readily applicable to other complex
interfacial and bulk systems. As a concrete example, Du \emph{et
al.}~\citep{du2024aluminumoxide} employed ai2-kit to compute the IR, Raman, and SFG
spectra of the aluminum oxide--water interface, achieving close agreement with
experimental observations and demonstrating the workflow's transferability across
chemically distinct environments.

\subsection{Example-driven workflow distribution}
\label{sec:examples}

As outlined in the section ``\nameref{sec:structure}'', ai2-kit exposes a wide family of command-line tools that
researchers can
compose into custom workflows for their own scientific problems. For users
who prefer to get started without first assembling these commands, ai2-kit also ships an
executable example for many applications, including TESLA-based MLP training, FEP,
ec-MLP, and spectroscopy, under \texttt{example/use-case/} in the ai2-kit GitHub
repository. Each example is a self-contained directory comprising
configuration templates, workflow scripts, and a launchable entry point (typically
\texttt{run.sh}), curated so that the entire pipeline runs end-to-end with minimal setup.
These examples serve as ready-to-modify templates that users can adapt to their own
systems through localized edits to the configuration files and a handful of parameters.

Furthermore, the concise and independent nature of these example-driven workflows makes
them exceptionally well-suited for integration with AI code agents. To assist users in
adapting workflows, ai2-kit provides AI agent skills in the \texttt{./skills/} directory.
For example, the \texttt{build-tesla} skill (a plugin mechanism designed to extend an AI
agent's capabilities) allows users to generate new workflows from existing TESLA examples
by simply substituting the underlying computational engines, such as the machine learning
potential, molecular dynamics, or \emph{ab initio} software.

For instance, we performed an experiment in which an AI code agent was
used to develop a new TESLA workflow. The workflow was initialized by copying the complete
\texttt{tesla-h2o} example into a new working directory named \texttt{skill-demo}, which
served as the starting point for the migration. The experiment was conducted in Visual
Studio Code using the GitHub Copilot code agent with the Claude Code Sonnet 4.6 model.
After installing the \texttt{build-tesla} skill, the following prompt was issued:

\begin{lstlisting}[style=shellstyle]
/build-tesla Build a TESLA workflow based on the current project, using MACE for the MLIP, OpenMM with the MACE potential for molecular dynamics exploration, and VASP for data labeling.
\end{lstlisting}

In response to \texttt{build-tesla}, the agent generated \path{iter-classic-mace-openmm-vasp.sh} as the top-level workflow script. It also populated the corresponding
configuration templates under \texttt{./00-config} for MACE, OpenMM, and VASP, wired the
OpenMM exploration stage to the MACE potential, and added the model-deviation utility
needed to grade candidate structures after each exploration run. This means that
users can start from the generated workflow, then continue refining cluster-specific
details by sending follow-up prompts to the agent or by editing the configuration files
manually, including software loading commands, job-submission parameters, MACE/OpenMM/VASP
settings, and active-learning thresholds. Once those project-specific details are in
place, the complete active-learning loop can be launched directly with
\texttt{./run.sh}.

In total, the agent modified or created 14 files. The core workflow file,
\path{01-workflow/iter-classic-mace-openmm-vasp.sh}, contains only 178 lines,
illustrating the compactness of workflow development based on ai2-kit. These files are
also available in the \texttt{example/} folder of the ai2-kit GitHub repository.

\begin{figure}[!htbp]
    \centering
    \includegraphics[width=0.8\linewidth]{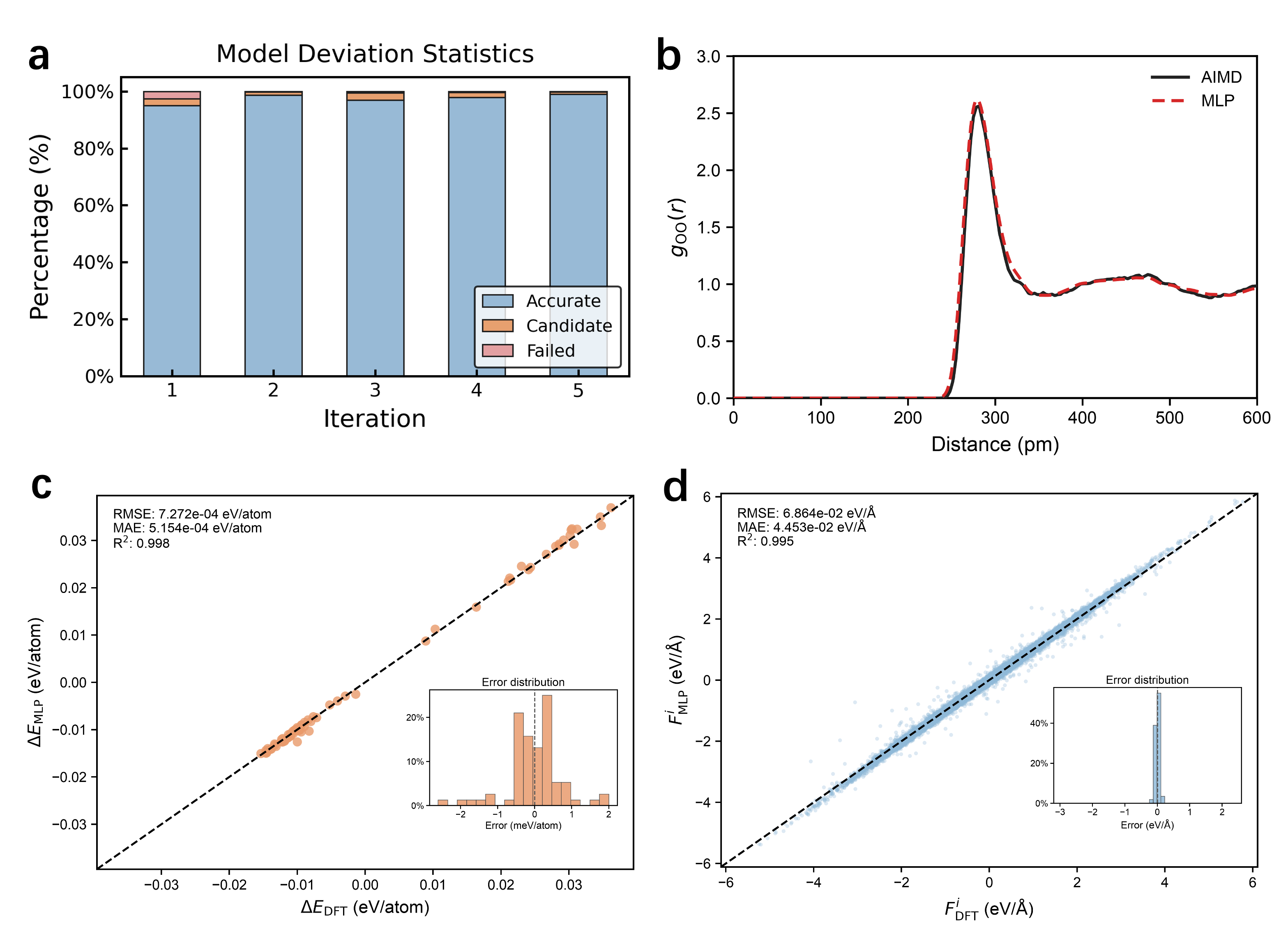}
    \caption{End-to-end MACE/OpenMM/VASP skill demonstration generated with
    \texttt{build-tesla}. (a) Model-deviation statistics across five active-learning
    iterations. (b) O--O radial distribution function (RDF) from MLMD compared with the
    reference AIMD result. (c,d) Validation accuracy of the trained MACE model, showing
    (c) energy and (d) force predictions against reference data.}
    \label{fig:tesla-skill-result}
\end{figure}

The figure summarizes the resulting workflow behavior. \Cref{fig:tesla-skill-result}a
shows that the generated workflow completes five active-learning iterations with most
structures remaining in the accurate region and only a small fraction routed to candidate
or failed bins for labeling or inspection. \Cref{fig:tesla-skill-result}b shows that the
MLMD trajectory reproduces the reference O--O radial distribution function, while
\Cref{fig:tesla-skill-result}c,d show tight energy and force agreement with the reference
labels. In the bulk-water test reported with this demo, the workflow collected 94 labeled
structures before reaching DeePMD-kit-comparable training accuracy, showing that ai2-kit
can use an AI agent to turn an example TESLA template into a working MACE/OpenMM/VASP
active-learning setup with only localized edits.

\section{Conclusions}
\label{sec:conclusion}

Complex chemical systems increasingly require simulations that combine electronic-structure
accuracy with extensive finite-temperature sampling. \AItwo{} methods provide a practical
route toward this goal, but their routine use in chemical research is often limited by the
difficulty of constructing reliable workflows across multiple software packages, data formats,
simulation engines, and high-performance-computing environments. In this work, we presented
ai2-kit as a toolkit designed to lower this practical barrier and make \AItwo{} workflow
development more accessible to chemists.

The examples demonstrated here show that ai2-kit is not limited to a single type of machine
learning potential or a single simulation protocol. Through its high-semantic-density
interfaces, composable command-line tools, and ready-to-modify workflow templates, ai2-kit
supports active-learning-based MLP construction, free-energy perturbation for redox and
acid--base processes, electrochemical machine learning potentials for electrified interfaces,
and MLMD-based spectroscopic prediction. These applications involve different chemical
questions and different combinations of training codes, molecular dynamics engines,
first-principles software, and analysis tools, yet they can be expressed within a common
workflow framework.

For users, the main benefit is that the effort of applying \AItwo{} methods can be shifted
from rebuilding technical infrastructure to formulating and testing chemical hypotheses.
Routine operations such as dataset conversion, batch task generation, model-deviation
screening, job submission, checkpoint recovery, and workflow adaptation are handled as
reusable components rather than repeatedly rewritten for each project. As a result,
researchers can more rapidly construct MLPs, extend them to specialized applications, and
validate the resulting simulations against thermodynamic, electrochemical, or spectroscopic
observables.

Looking forward, we envision ai2-kit as an example-driven and extensible ecosystem for
\AItwo{} chemistry. New workflow templates can be developed from existing examples by
changing the target system, computational engine, sampling strategy, or property model,
while preserving the same underlying workflow logic. By making these workflows easier to
build, modify, reproduce, and share, ai2-kit can help accelerate the adoption of
AI-accelerated ab initio simulation in the study of complex chemical systems.

\section{Author contributions}
\label{sec:author-contributions}

\textbf{Conceptualization}: J.C.
\textbf{Methodology}: S.B., X.Y., F.W., F.T., W.X., J.C.
\textbf{Software}: All authors.
\textbf{Writing}: All authors.
\textbf{Supervision}: J.C.

\section{Conflicts of interest}
\label{sec:coi}

There are no conflicts to declare.

\section{Data availability}
\label{sec:data-availability}

ai2-kit is open source and available on GitHub at
\url{https://github.com/chenggroup/ai2-kit}, together with its documentation
and example workflows. All cases demonstrated in this paper are provided as
ready-to-run use cases in the repository and can be reproduced from there.

\section{Acknowledgements}
\label{sec:acknowledgements}

This work was supported by the National Natural Science Foundation of China (Grant
Nos.~22225302, 92470201, 92461312, 22021001, 21991151, 21991150, 92161113, and
22411560277 to J.C.; Grant Nos.~22393901 and 22503037 to F.W.; Grant No.~22503072 to
S.B.), the New Generation Artificial Intelligence -- National Science and Technology
Major Project (Grant No.~2025ZD0122703), the Fundamental Research Funds for the Central
Universities (Grant Nos.~20720250005, 20720220009, and 20720230090 to J.C.), and the
Laboratory of AI for Electrochemistry (AI4EC) and IKKEM (Grant Nos.~RD2023100101 and
RD2022070501 to J.C. and F.W.).

\section{Notes and references}
\label{sec:references}

\bibliographystyle{rsc-titlelink}
\bibliography{draft}

\providecommand*{\mcitethebibliography}{\thebibliography}
\csname @ifundefined\endcsname{endmcitethebibliography}
{\let\endmcitethebibliography\endthebibliography}{}
\begin{mcitethebibliography}{60}
\providecommand*{\natexlab}[1]{#1}
\providecommand*{\mciteSetBstSublistMode}[1]{}
\providecommand*{\mciteSetBstMaxWidthForm}[2]{}
\providecommand*{\mciteBstWouldAddEndPuncttrue}
  {\def\EndOfBibitem{\unskip.}}
\providecommand*{\mciteBstWouldAddEndPunctfalse}
  {\let\EndOfBibitem\relax}
\providecommand*{\mciteSetBstMidEndSepPunct}[3]{}
\providecommand*{\mciteSetBstSublistLabelBeginEnd}[3]{}
\providecommand*{\EndOfBibitem}{}
\mciteSetBstSublistMode{f}
\mciteSetBstMaxWidthForm{subitem}
{(\emph{\alph{mcitesubitemcount}})}
\mciteSetBstSublistLabelBeginEnd{\mcitemaxwidthsubitemform\space}
{\relax}{\relax}

\bibitem[Truhlar(2008)]{truhlar2008complex}
D.~G. Truhlar, \href{https://doi.org/10.1021/ja808927h}{Molecular Modeling of Complex Chemical Systems}, \emph{Journal of the American Chemical Society}, 2008, \textbf{130}, 16824--16827, \href{https://doi.org/10.1021/ja808927h}{DOI: 10.1021/ja808927h}\relax
\mciteBstWouldAddEndPuncttrue
\mciteSetBstMidEndSepPunct{\mcitedefaultmidpunct}
{\mcitedefaultendpunct}{\mcitedefaultseppunct}\relax
\EndOfBibitem
\bibitem[Gastegger \emph{et~al.}(2021)Gastegger, Sch{\"u}tt, and M{\"u}ller]{gastegger2021solvent}
M.~Gastegger, K.~T. Sch{\"u}tt and K.-R. M{\"u}ller, \href{https://doi.org/10.1039/D1SC02742E}{Machine learning of solvent effects on molecular spectra and reactions}, \emph{Chemical Science}, 2021, \textbf{12}, 11473--11483, \href{https://doi.org/10.1039/D1SC02742E}{DOI: 10.1039/D1SC02742E}\relax
\mciteBstWouldAddEndPuncttrue
\mciteSetBstMidEndSepPunct{\mcitedefaultmidpunct}
{\mcitedefaultendpunct}{\mcitedefaultseppunct}\relax
\EndOfBibitem
\bibitem[Goldsmith \emph{et~al.}(2018)Goldsmith, Esterhuizen, Liu, Bartel, and Sutton]{goldsmith2018catalysis}
B.~R. Goldsmith, J.~Esterhuizen, J.-X. Liu, C.~J. Bartel and C.~Sutton, \href{https://doi.org/10.1002/aic.16198}{Machine learning for heterogeneous catalyst design and discovery}, \emph{AIChE Journal}, 2018, \textbf{64}, 2311--2323, \href{https://doi.org/10.1002/aic.16198}{DOI: 10.1002/aic.16198}\relax
\mciteBstWouldAddEndPuncttrue
\mciteSetBstMidEndSepPunct{\mcitedefaultmidpunct}
{\mcitedefaultendpunct}{\mcitedefaultseppunct}\relax
\EndOfBibitem
\bibitem[Xue \emph{et~al.}(2024)Xue, Chaudhary, Nouri, Gubanova, Garlyyev, Alexandrov, and Bandarenka]{pt2024doublelayer}
S.~Xue, P.~Chaudhary, M.~R. Nouri, E.~Gubanova, B.~Garlyyev, V.~Alexandrov and A.~S. Bandarenka, \href{https://doi.org/10.1021/jacs.3c11403}{Impact of {Pt}(hkl) Electrode Surface Structure on the Electrical Double Layer Capacitance}, \emph{Journal of the American Chemical Society}, 2024, \textbf{146}, 3883--3889, \href{https://doi.org/10.1021/jacs.3c11403}{DOI: 10.1021/jacs.3c11403}\relax
\mciteBstWouldAddEndPuncttrue
\mciteSetBstMidEndSepPunct{\mcitedefaultmidpunct}
{\mcitedefaultendpunct}{\mcitedefaultseppunct}\relax
\EndOfBibitem
\bibitem[Wang \emph{et~al.}(2024)Wang, Ma, and Cheng]{wang2024redox}
F.~Wang, Z.~Ma and J.~Cheng, \href{https://doi.org/10.1021/jacs.4c01221}{Accelerating Computation of Acidity Constants and Redox Potentials for Aqueous Organic Redox Flow Batteries by Machine Learning Potential-Based Molecular Dynamics}, \emph{Journal of the American Chemical Society}, 2024, \textbf{146}, 14566--14575, \href{https://doi.org/10.1021/jacs.4c01221}{DOI: 10.1021/jacs.4c01221}\relax
\mciteBstWouldAddEndPuncttrue
\mciteSetBstMidEndSepPunct{\mcitedefaultmidpunct}
{\mcitedefaultendpunct}{\mcitedefaultseppunct}\relax
\EndOfBibitem
\bibitem[Cheng \emph{et~al.}(2021)Cheng, Bethkenhagen, Pickard, and Hamel]{cheng2021water}
B.~Cheng, M.~Bethkenhagen, C.~J. Pickard and S.~Hamel, \href{https://doi.org/10.1038/s41567-021-01334-9}{Phase behaviours of superionic water at planetary conditions}, \emph{Nature Physics}, 2021, \textbf{17}, 1228--1232, \href{https://doi.org/10.1038/s41567-021-01334-9}{DOI: 10.1038/s41567-021-01334-9}\relax
\mciteBstWouldAddEndPuncttrue
\mciteSetBstMidEndSepPunct{\mcitedefaultmidpunct}
{\mcitedefaultendpunct}{\mcitedefaultseppunct}\relax
\EndOfBibitem
\bibitem[Wang and Cheng(2022)]{wang2022electrolyte}
F.~Wang and J.~Cheng, \href{https://doi.org/10.1039/D2SC04025E}{Unraveling the origin of reductive stability of super-concentrated electrolytes from first principles and unsupervised machine learning}, \emph{Chemical Science}, 2022, \textbf{13}, 11570--11576, \href{https://doi.org/10.1039/D2SC04025E}{DOI: 10.1039/D2SC04025E}\relax
\mciteBstWouldAddEndPuncttrue
\mciteSetBstMidEndSepPunct{\mcitedefaultmidpunct}
{\mcitedefaultendpunct}{\mcitedefaultseppunct}\relax
\EndOfBibitem
\bibitem[Wang \emph{et~al.}(2023)Wang, Sun, and Cheng]{wang2023redox}
F.~Wang, Y.~Sun and J.~Cheng, \href{https://doi.org/10.1021/jacs.2c11793}{Switching of Redox Levels Leads to High Reductive Stability in Water-in-Salt Electrolytes}, \emph{Journal of the American Chemical Society}, 2023, \textbf{145}, 4056--4064, \href{https://doi.org/10.1021/jacs.2c11793}{DOI: 10.1021/jacs.2c11793}\relax
\mciteBstWouldAddEndPuncttrue
\mciteSetBstMidEndSepPunct{\mcitedefaultmidpunct}
{\mcitedefaultendpunct}{\mcitedefaultseppunct}\relax
\EndOfBibitem
\bibitem[Jia \emph{et~al.}(2024)Jia, Zhuang, Wang, Zhang, and Cheng]{sno2proton2024}
M.~Jia, Y.-B. Zhuang, F.~Wang, C.~Zhang and J.~Cheng, \href{https://doi.org/10.1021/prechem.4c00056}{Water-Mediated Proton Hopping Mechanisms at the {SnO$_2$(110)/H$_2$O} Interface from Ab Initio Deep Potential Molecular Dynamics}, \emph{Precision Chemistry}, 2024, \textbf{2}, 644--654, \href{https://doi.org/10.1021/prechem.4c00056}{DOI: 10.1021/prechem.4c00056}\relax
\mciteBstWouldAddEndPuncttrue
\mciteSetBstMidEndSepPunct{\mcitedefaultmidpunct}
{\mcitedefaultendpunct}{\mcitedefaultseppunct}\relax
\EndOfBibitem
\bibitem[Kornyshev(2007)]{kornyshev2007double}
A.~A. Kornyshev, \href{https://doi.org/10.1021/jp067857o}{Double-Layer in Ionic Liquids: Paradigm Change?}, \emph{The Journal of Physical Chemistry B}, 2007, \textbf{111}, 5545--5557, \href{https://doi.org/10.1021/jp067857o}{DOI: 10.1021/jp067857o}\relax
\mciteBstWouldAddEndPuncttrue
\mciteSetBstMidEndSepPunct{\mcitedefaultmidpunct}
{\mcitedefaultendpunct}{\mcitedefaultseppunct}\relax
\EndOfBibitem
\bibitem[Bazant \emph{et~al.}(2011)Bazant, Storey, and Kornyshev]{bazant2011overscreening}
M.~Z. Bazant, B.~D. Storey and A.~A. Kornyshev, \href{https://doi.org/10.1103/PhysRevLett.106.046102}{Double Layer in Ionic Liquids: Overscreening versus Crowding}, \emph{Physical Review Letters}, 2011, \textbf{106}, 046102, \href{https://doi.org/10.1103/PhysRevLett.106.046102}{DOI: 10.1103/PhysRevLett.106.046102}\relax
\mciteBstWouldAddEndPuncttrue
\mciteSetBstMidEndSepPunct{\mcitedefaultmidpunct}
{\mcitedefaultendpunct}{\mcitedefaultseppunct}\relax
\EndOfBibitem
\bibitem[Tang \emph{et~al.}(2020)Tang, Ohto, Sun, Rouxel, Imoto, Backus, Mukamel, Bonn, and Nagata]{tang2020molecular}
F.~Tang, T.~Ohto, S.~Sun, J.~R. Rouxel, S.~Imoto, E.~H.~G. Backus, S.~Mukamel, M.~Bonn and Y.~Nagata, \href{https://doi.org/10.1021/acs.chemrev.9b00512}{Molecular Structure and Modeling of Water--Air and Ice--Air Interfaces Monitored by Sum-Frequency Generation}, \emph{Chemical Reviews}, 2020, \textbf{120}, 3633--3667, \href{https://doi.org/10.1021/acs.chemrev.9b00512}{DOI: 10.1021/acs.chemrev.9b00512}\relax
\mciteBstWouldAddEndPuncttrue
\mciteSetBstMidEndSepPunct{\mcitedefaultmidpunct}
{\mcitedefaultendpunct}{\mcitedefaultseppunct}\relax
\EndOfBibitem
\bibitem[Kaliannan \emph{et~al.}(2020)Kaliannan, Henao~Aristizabal, Wiebeler, Zysk, Ohto, Nagata, and K{\"u}hne]{ohto2019coupling}
N.~K. Kaliannan, A.~Henao~Aristizabal, H.~Wiebeler, F.~Zysk, T.~Ohto, Y.~Nagata and T.~D. K{\"u}hne, \href{https://doi.org/10.1080/00268976.2019.1620358}{Impact of intermolecular vibrational coupling effects on the sum-frequency generation spectra of the water/air interface}, \emph{Molecular Physics}, 2020, \textbf{118}, 1620358, \href{https://doi.org/10.1080/00268976.2019.1620358}{DOI: 10.1080/00268976.2019.1620358}\relax
\mciteBstWouldAddEndPuncttrue
\mciteSetBstMidEndSepPunct{\mcitedefaultmidpunct}
{\mcitedefaultendpunct}{\mcitedefaultseppunct}\relax
\EndOfBibitem
\bibitem[Seki \emph{et~al.}(2021)Seki, Yu, Chiang, Tan, Sun, Ye, Bonn, and Nagata]{yu2021bending}
T.~Seki, C.-C. Yu, K.-Y. Chiang, J.~Tan, S.~Sun, S.~Ye, M.~Bonn and Y.~Nagata, \href{https://doi.org/10.1021/acs.jpcb.1c03258}{Disentangling Sum-Frequency Generation Spectra of the Water Bending Mode at Charged Aqueous Interfaces}, \emph{The Journal of Physical Chemistry B}, 2021, \textbf{125}, 7060--7067, \href{https://doi.org/10.1021/acs.jpcb.1c03258}{DOI: 10.1021/acs.jpcb.1c03258}\relax
\mciteBstWouldAddEndPuncttrue
\mciteSetBstMidEndSepPunct{\mcitedefaultmidpunct}
{\mcitedefaultendpunct}{\mcitedefaultseppunct}\relax
\EndOfBibitem
\bibitem[Du \emph{et~al.}(2024)Du, Shao, Bao, Zhang, Cheng, and Tang]{du2024aluminumoxide}
X.~Du, W.~Shao, C.~Bao, L.~Zhang, J.~Cheng and F.~Tang, \href{https://doi.org/10.1063/5.0230101}{Revealing the molecular structures of {$\alpha$-Al$_2$O$_3$}(0001)--water interface by machine learning based computational vibrational spectroscopy}, \emph{The Journal of Chemical Physics}, 2024, \textbf{161}, 124702, \href{https://doi.org/10.1063/5.0230101}{DOI: 10.1063/5.0230101}\relax
\mciteBstWouldAddEndPuncttrue
\mciteSetBstMidEndSepPunct{\mcitedefaultmidpunct}
{\mcitedefaultendpunct}{\mcitedefaultseppunct}\relax
\EndOfBibitem
\bibitem[Marx and Hutter(2009)]{marx2009aimd}
D.~Marx and J.~Hutter, \emph{\href{https://doi.org/10.1017/CBO9780511609633}{Ab Initio Molecular Dynamics: Basic Theory and Advanced Methods}}, Cambridge University Press, 2009, \href{https://doi.org/10.1017/CBO9780511609633}{DOI: 10.1017/CBO9780511609633}\relax
\mciteBstWouldAddEndPuncttrue
\mciteSetBstMidEndSepPunct{\mcitedefaultmidpunct}
{\mcitedefaultendpunct}{\mcitedefaultseppunct}\relax
\EndOfBibitem
\bibitem[K{\"u}hne(2014)]{kuhne2014firstprinciples}
T.~D. K{\"u}hne, \href{https://doi.org/10.1002/wcms.1176}{Second generation Car--Parrinello molecular dynamics}, \emph{WIREs Computational Molecular Science}, 2014, \textbf{4}, 391--406, \href{https://doi.org/10.1002/wcms.1176}{DOI: 10.1002/wcms.1176}\relax
\mciteBstWouldAddEndPuncttrue
\mciteSetBstMidEndSepPunct{\mcitedefaultmidpunct}
{\mcitedefaultendpunct}{\mcitedefaultseppunct}\relax
\EndOfBibitem
\bibitem[Jia \emph{et~al.}(2020)Jia, Wang, Chen, Lu, Lin, Car, E, and Zhang]{jia2020pushing}
W.~Jia, H.~Wang, M.~Chen, D.~Lu, L.~Lin, R.~Car, W.~E and L.~Zhang, \href{https://doi.org/10.1109/SC41405.2020.00009}{Pushing the Limit of Molecular Dynamics with Ab Initio Accuracy to 100 Million Atoms with Machine Learning}, \href{https://doi.org/10.1109/SC41405.2020.00009}{{SC20}: International Conference for High Performance Computing, Networking, Storage and Analysis}, 2020, pp. 1--14, \href{https://doi.org/10.1109/SC41405.2020.00009}{DOI: 10.1109/SC41405.2020.00009}\relax
\mciteBstWouldAddEndPuncttrue
\mciteSetBstMidEndSepPunct{\mcitedefaultmidpunct}
{\mcitedefaultendpunct}{\mcitedefaultseppunct}\relax
\EndOfBibitem
\bibitem[Behler(2021)]{behler2021fourth}
J.~Behler, \href{https://doi.org/10.1021/acs.chemrev.0c00868}{Four Generations of High-Dimensional Neural Network Potentials}, \emph{Chemical Reviews}, 2021, \textbf{121}, 10037--10072, \href{https://doi.org/10.1021/acs.chemrev.0c00868}{DOI: 10.1021/acs.chemrev.0c00868}\relax
\mciteBstWouldAddEndPuncttrue
\mciteSetBstMidEndSepPunct{\mcitedefaultmidpunct}
{\mcitedefaultendpunct}{\mcitedefaultseppunct}\relax
\EndOfBibitem
\bibitem[Unke \emph{et~al.}(2021)Unke, Chmiela, Sauceda, Gastegger, Poltavsky, Sch{\"u}tt, Tkatchenko, and M{\"u}ller]{unke2021machine}
O.~T. Unke, S.~Chmiela, H.~E. Sauceda, M.~Gastegger, I.~Poltavsky, K.~T. Sch{\"u}tt, A.~Tkatchenko and K.-R. M{\"u}ller, \href{https://doi.org/10.1021/acs.chemrev.0c01111}{Machine Learning Force Fields}, \emph{Chemical Reviews}, 2021, \textbf{121}, 10142--10186, \href{https://doi.org/10.1021/acs.chemrev.0c01111}{DOI: 10.1021/acs.chemrev.0c01111}\relax
\mciteBstWouldAddEndPuncttrue
\mciteSetBstMidEndSepPunct{\mcitedefaultmidpunct}
{\mcitedefaultendpunct}{\mcitedefaultseppunct}\relax
\EndOfBibitem
\bibitem[Deringer \emph{et~al.}(2019)Deringer, Caro, and Cs{\'a}nyi]{deringer2021gaussian}
V.~L. Deringer, M.~A. Caro and G.~Cs{\'a}nyi, \href{https://doi.org/10.1002/adma.201902765}{Machine Learning Interatomic Potentials as Emerging Tools for Materials Science}, \emph{Advanced Materials}, 2019, \textbf{31}, 1902765, \href{https://doi.org/10.1002/adma.201902765}{DOI: 10.1002/adma.201902765}\relax
\mciteBstWouldAddEndPuncttrue
\mciteSetBstMidEndSepPunct{\mcitedefaultmidpunct}
{\mcitedefaultendpunct}{\mcitedefaultseppunct}\relax
\EndOfBibitem
\bibitem[Fan \emph{et~al.}(2024)Fan, Liu, Zhu, Gong, Wang, E, Bao, Tian, and Cheng]{D4SC05399K}
Q.-Y. Fan, Y.-P. Liu, H.-X. Zhu, F.-Q. Gong, Y.~Wang, W.~E, X.~Bao, Z.-Q. Tian and J.~Cheng, \href{https://doi.org/10.1039/D4SC05399K}{Entropy in catalyst dynamics under confinement}, \emph{Chemical Science}, 2024, \textbf{15}, 18303--18309, \href{https://doi.org/10.1039/D4SC05399K}{DOI: 10.1039/D4SC05399K}\relax
\mciteBstWouldAddEndPuncttrue
\mciteSetBstMidEndSepPunct{\mcitedefaultmidpunct}
{\mcitedefaultendpunct}{\mcitedefaultseppunct}\relax
\EndOfBibitem
\bibitem[Gong \emph{et~al.}(2024)Gong, Liu, Wang, E, Tian, and Cheng]{gong2024anie}
F.-Q. Gong, Y.-P. Liu, Y.~Wang, W.~E, Z.-Q. Tian and J.~Cheng, \href{https://doi.org/10.1002/anie.202405379}{Machine Learning Molecular Dynamics Shows Anomalous Entropic Effect on Catalysis through Surface Pre-melting of Nanoclusters}, \emph{Angewandte Chemie International Edition}, 2024, \textbf{63}, e202405379, \href{https://doi.org/10.1002/anie.202405379}{DOI: 10.1002/anie.202405379}\relax
\mciteBstWouldAddEndPuncttrue
\mciteSetBstMidEndSepPunct{\mcitedefaultmidpunct}
{\mcitedefaultendpunct}{\mcitedefaultseppunct}\relax
\EndOfBibitem
\bibitem[Zhang \emph{et~al.}(2024)Zhang, Chen, and Liu]{zhang2024jacs}
K.-X. Zhang, L.~Chen and Z.-P. Liu, \href{https://doi.org/10.1021/jacs.4c14404}{Do Rh-Hydride Phases Contribute to the Catalytic Activity of Rh Catalysts under Reductive Conditions?}, \emph{Journal of the American Chemical Society}, 2024, \textbf{146}, 35416--35426, \href{https://doi.org/10.1021/jacs.4c14404}{DOI: 10.1021/jacs.4c14404}\relax
\mciteBstWouldAddEndPuncttrue
\mciteSetBstMidEndSepPunct{\mcitedefaultmidpunct}
{\mcitedefaultendpunct}{\mcitedefaultseppunct}\relax
\EndOfBibitem
\bibitem[Zhou \emph{et~al.}(2024)Zhou, Fu, Wang, Wang, Luo, Yan, He, Jia, Li, and Liu]{zhou2024sciadv}
L.~Zhou, X.-P. Fu, R.~Wang, C.-X. Wang, F.~Luo, H.~Yan, Y.~He, C.-J. Jia, J.~Li and J.-C. Liu, \href{https://doi.org/10.1126/sciadv.adr4145}{Dynamic phase transitions dictate the size effect and activity of supported gold catalysts}, \emph{Science Advances}, 2024, \textbf{10}, eadr4145, \href{https://doi.org/10.1126/sciadv.adr4145}{DOI: 10.1126/sciadv.adr4145}\relax
\mciteBstWouldAddEndPuncttrue
\mciteSetBstMidEndSepPunct{\mcitedefaultmidpunct}
{\mcitedefaultendpunct}{\mcitedefaultseppunct}\relax
\EndOfBibitem
\bibitem[Zhang \emph{et~al.}(2020)Zhang, Wang, Chen, Zeng, Zhang, Wang, and E]{zhang2020dpgen}
Y.~Zhang, H.~Wang, W.~Chen, J.~Zeng, L.~Zhang, H.~Wang and W.~E, \href{https://doi.org/10.1016/j.cpc.2020.107206}{{DP-GEN}: A concurrent learning platform for the generation of reliable deep learning based potential energy models}, \emph{Computer Physics Communications}, 2020, \textbf{253}, 107206, \href{https://doi.org/10.1016/j.cpc.2020.107206}{DOI: 10.1016/j.cpc.2020.107206}\relax
\mciteBstWouldAddEndPuncttrue
\mciteSetBstMidEndSepPunct{\mcitedefaultmidpunct}
{\mcitedefaultendpunct}{\mcitedefaultseppunct}\relax
\EndOfBibitem
\bibitem[Vandermause \emph{et~al.}(2020)Vandermause, Torrisi, Batzner, Xie, Sun, Kolpak, and Kozinsky]{vandermause2020flare}
J.~Vandermause, S.~B. Torrisi, S.~Batzner, Y.~Xie, L.~Sun, A.~M. Kolpak and B.~Kozinsky, \href{https://doi.org/10.1038/s41524-020-0283-z}{On-the-fly active learning of interpretable Bayesian force fields for atomistic rare events}, \emph{npj Computational Materials}, 2020, \textbf{6}, 20, \href{https://doi.org/10.1038/s41524-020-0283-z}{DOI: 10.1038/s41524-020-0283-z}\relax
\mciteBstWouldAddEndPuncttrue
\mciteSetBstMidEndSepPunct{\mcitedefaultmidpunct}
{\mcitedefaultendpunct}{\mcitedefaultseppunct}\relax
\EndOfBibitem
\bibitem[Babuji \emph{et~al.}(2019)Babuji, Woodard, Li, Katz, Clifford, Kumar, Lacinski, Chard, Wozniak, Foster, Wilde, and Chard]{babuji2019parsl}
Y.~Babuji, A.~Woodard, Z.~Li, D.~S. Katz, B.~Clifford, R.~Kumar, L.~Lacinski, R.~Chard, J.~M. Wozniak, I.~Foster, M.~Wilde and K.~Chard, \href{https://doi.org/10.1145/3307681.3325400}{Parsl: Pervasive Parallel Programming in Python}, \href{https://doi.org/10.1145/3307681.3325400}{Proceedings of the 28th International Symposium on High-Performance Parallel and Distributed Computing}, 2019, pp. 25--36, \href{https://doi.org/10.1145/3307681.3325400}{DOI: 10.1145/3307681.3325400}\relax
\mciteBstWouldAddEndPuncttrue
\mciteSetBstMidEndSepPunct{\mcitedefaultmidpunct}
{\mcitedefaultendpunct}{\mcitedefaultseppunct}\relax
\EndOfBibitem
\bibitem[Jain \emph{et~al.}(2015)Jain, Ong, Chen, Medasani, Qu, Kocher, Brafman, Petretto, Rignanese, Hautier, Gunter, and Persson]{jain2015fireworks}
A.~Jain, S.~P. Ong, W.~Chen, B.~Medasani, X.~Qu, M.~Kocher, M.~Brafman, G.~Petretto, G.-M. Rignanese, G.~Hautier, D.~Gunter and K.~A. Persson, \href{https://doi.org/10.1002/cpe.3505}{FireWorks: a dynamic workflow system designed for high-throughput applications}, \emph{Concurrency and Computation: Practice and Experience}, 2015, \textbf{27}, 5037--5059, \href{https://doi.org/10.1002/cpe.3505}{DOI: 10.1002/cpe.3505}\relax
\mciteBstWouldAddEndPuncttrue
\mciteSetBstMidEndSepPunct{\mcitedefaultmidpunct}
{\mcitedefaultendpunct}{\mcitedefaultseppunct}\relax
\EndOfBibitem
\bibitem[Liu \emph{et~al.}(2024)Liu, Han, Li, Fan, Zhang, Zeng, Shan, Yuan, Xu, Liu, Zhang, Wen, York, Zhong, Zheng, Cheng, Zhang, and Wang]{dflow2024}
X.~Liu, Y.~Han, Z.~Li, J.~Fan, C.~Zhang, J.~Zeng, Y.~Shan, Y.~Yuan, W.-H. Xu, Y.-P. Liu, Y.~Zhang, T.~Wen, D.~M. York, Z.~Zhong, H.~Zheng, J.~Cheng, L.~Zhang and H.~Wang, \emph{\href{https://doi.org/10.48550/arXiv.2404.18392}{Dflow, a Python framework for constructing cloud-native AI-for-Science workflows}}, arXiv:2404.18392, 2024, \url{https://arxiv.org/abs/2404.18392}, \href{https://doi.org/10.48550/arXiv.2404.18392}{DOI: 10.48550/arXiv.2404.18392}\relax
\mciteBstWouldAddEndPuncttrue
\mciteSetBstMidEndSepPunct{\mcitedefaultmidpunct}
{\mcitedefaultendpunct}{\mcitedefaultseppunct}\relax
\EndOfBibitem
\bibitem[M{\"o}lder \emph{et~al.}(2021)M{\"o}lder, Jablonski, Letcher, Hall, Tomkins-Tinch, Sochat, Forster, Lee, Twardziok, Kanitz, Wilm, Holtgrewe, Rahmann, Nahnsen, and K{\"o}ster]{molder2021snakemake}
F.~M{\"o}lder, K.~P. Jablonski, B.~Letcher, M.~B. Hall, C.~H. Tomkins-Tinch, V.~Sochat, J.~Forster, S.~Lee, S.~O. Twardziok, A.~Kanitz, A.~Wilm, M.~Holtgrewe, S.~Rahmann, S.~Nahnsen and J.~K{\"o}ster, \href{https://doi.org/10.12688/f1000research.29032.2}{Sustainable data analysis with Snakemake}, \emph{F1000Research}, 2021, \textbf{10}, 33, \href{https://doi.org/10.12688/f1000research.29032.2}{DOI: 10.12688/f1000research.29032.2}\relax
\mciteBstWouldAddEndPuncttrue
\mciteSetBstMidEndSepPunct{\mcitedefaultmidpunct}
{\mcitedefaultendpunct}{\mcitedefaultseppunct}\relax
\EndOfBibitem
\bibitem[K{\"u}hne \emph{et~al.}(2020)K{\"u}hne, Iannuzzi, Del~Ben, Rybkin, Seewald, Stein, Laino, Khaliullin, Sch{\"u}tt, Schiffmann, Golze, Wilhelm, Chulkov, Bani-Hashemian, Weber, Bor{\v{s}}tnik, Taillefumier, Jakobovits, Lazzaro, Pabst, M{\"u}ller, Schade, Guidon, Andermatt, Holmberg, Schenter, Hehn, Bussy, Belleflamme, Tabacchi, Gl{\"o}{\ss}, Lass, Bethune, Mundy, Plessl, Watkins, VandeVondele, Krack, and Hutter]{kuhne2020cp2k}
T.~D. K{\"u}hne, M.~Iannuzzi, M.~Del~Ben, V.~V. Rybkin, P.~Seewald, F.~Stein, T.~Laino, R.~Z. Khaliullin, O.~Sch{\"u}tt, F.~Schiffmann, D.~Golze, J.~Wilhelm, S.~Chulkov, M.~H. Bani-Hashemian, V.~Weber, U.~Bor{\v{s}}tnik, M.~Taillefumier, A.~S. Jakobovits, A.~Lazzaro, H.~Pabst, T.~M{\"u}ller, R.~Schade, M.~Guidon, S.~Andermatt, N.~Holmberg, G.~K. Schenter, A.~Hehn, A.~Bussy, F.~Belleflamme, G.~Tabacchi, A.~Gl{\"o}{\ss}, M.~Lass, I.~Bethune, C.~J. Mundy, C.~Plessl, M.~Watkins, J.~VandeVondele, M.~Krack and J.~Hutter, \href{https://doi.org/10.1063/5.0007045}{{CP2K}: An electronic structure and molecular dynamics software package---Quickstep: Efficient and accurate electronic structure calculations}, \emph{The Journal of Chemical Physics}, 2020, \textbf{152}, 194103, \href{https://doi.org/10.1063/5.0007045}{DOI: 10.1063/5.0007045}\relax
\mciteBstWouldAddEndPuncttrue
\mciteSetBstMidEndSepPunct{\mcitedefaultmidpunct}
{\mcitedefaultendpunct}{\mcitedefaultseppunct}\relax
\EndOfBibitem
\bibitem[Kresse and Furthm{\"u}ller(1996)]{kresse1996vasp}
G.~Kresse and J.~Furthm{\"u}ller, \href{https://doi.org/10.1103/PhysRevB.54.11169}{Efficient iterative schemes for ab initio total-energy calculations using a plane-wave basis set}, \emph{Physical Review B}, 1996, \textbf{54}, 11169--11186, \href{https://doi.org/10.1103/PhysRevB.54.11169}{DOI: 10.1103/PhysRevB.54.11169}\relax
\mciteBstWouldAddEndPuncttrue
\mciteSetBstMidEndSepPunct{\mcitedefaultmidpunct}
{\mcitedefaultendpunct}{\mcitedefaultseppunct}\relax
\EndOfBibitem
\bibitem[Zhou \emph{et~al.}(2025)Zhou, Zheng, Liu, Lu, Liu, Lin, Huang, Peng, Bao, Cai, Jin, Wu, Zhang, Jin, Ji, Shen, Liu, Sun, Cao, Sun, Liu, Chen, Liu, Li, Han, Liang, Bao, Deng, Liu, Chen, Ren, Zhang, Liu, Fu, Liu, Li, Wen, Tang, Xu, Duan, Wang, Gu, Dai, Zheng, Zhong, Xiang, Gong, Zhao, Zhang, Ou, Jiang, Liu, Xu, Xu, Ren, He, Zhang, and Chen]{zhou2025abacus}
W.~Zhou, D.~Zheng, Q.~Liu, D.~Lu, Y.~Liu, P.~Lin, Y.~Huang, X.~Peng, J.~J. Bao, C.~Cai, Z.~Jin, J.~Wu, H.~Zhang, G.~Jin, Y.~Ji, Z.~Shen, X.~Liu, L.~Sun, Y.~Cao, M.~Sun, J.~Liu, T.~Chen, R.~Liu, Y.~Li, H.~Han, X.~Liang, T.~Bao, Z.~Deng, T.~Liu, N.~Chen, H.~Ren, X.~Zhang, Z.~Liu, Y.~Fu, M.~Liu, Z.~Li, T.~Wen, Z.~Tang, Y.~Xu, W.~Duan, X.~Wang, Q.~Gu, F.-Z. Dai, Q.~Zheng, Y.~Zhong, H.~Xiang, X.~Gong, J.~Zhao, Y.~Zhang, Q.~Ou, H.~Jiang, S.~Liu, B.~Xu, S.~Xu, X.~Ren, L.~He, L.~Zhang and M.~Chen, \href{https://doi.org/10.1063/5.0297563}{{ABACUS}: An electronic structure analysis package for the {AI} era}, \emph{The Journal of Chemical Physics}, 2025, \textbf{163}, 192501, \href{https://doi.org/10.1063/5.0297563}{DOI: 10.1063/5.0297563}\relax
\mciteBstWouldAddEndPuncttrue
\mciteSetBstMidEndSepPunct{\mcitedefaultmidpunct}
{\mcitedefaultendpunct}{\mcitedefaultseppunct}\relax
\EndOfBibitem
\bibitem[Plimpton(1995)]{plimpton1995lammps}
S.~Plimpton, \href{https://doi.org/10.1006/jcph.1995.1039}{Fast Parallel Algorithms for Short-Range Molecular Dynamics}, \emph{Journal of Computational Physics}, 1995, \textbf{117}, 1--19, \href{https://doi.org/10.1006/jcph.1995.1039}{DOI: 10.1006/jcph.1995.1039}\relax
\mciteBstWouldAddEndPuncttrue
\mciteSetBstMidEndSepPunct{\mcitedefaultmidpunct}
{\mcitedefaultendpunct}{\mcitedefaultseppunct}\relax
\EndOfBibitem
\bibitem[Bonomi \emph{et~al.}(2009)Bonomi, Branduardi, Bussi, Camilloni, Provasi, Raiteri, Donadio, Marinelli, Pietrucci, Broglia, and Parrinello]{bonomi2009plumed}
M.~Bonomi, D.~Branduardi, G.~Bussi, C.~Camilloni, D.~Provasi, P.~Raiteri, D.~Donadio, F.~Marinelli, F.~Pietrucci, R.~A. Broglia and M.~Parrinello, \href{https://doi.org/10.1016/j.cpc.2009.05.011}{{PLUMED}: A portable plugin for free-energy calculations with molecular dynamics}, \emph{Computer Physics Communications}, 2009, \textbf{180}, 1961--1972, \href{https://doi.org/10.1016/j.cpc.2009.05.011}{DOI: 10.1016/j.cpc.2009.05.011}\relax
\mciteBstWouldAddEndPuncttrue
\mciteSetBstMidEndSepPunct{\mcitedefaultmidpunct}
{\mcitedefaultendpunct}{\mcitedefaultseppunct}\relax
\EndOfBibitem
\bibitem[Schoenholz and Cubuk(2020)]{jaxmd2020}
S.~S. Schoenholz and E.~D. Cubuk, \href{https://papers.nips.cc/paper/2020/file/83d3d4b6c9579515e1679aca8cbc8033-Paper.pdf}{JAX M.D. A Framework for Differentiable Physics}, \href{https://papers.nips.cc/paper/2020/file/83d3d4b6c9579515e1679aca8cbc8033-Paper.pdf}{Advances in Neural Information Processing Systems}, 2020\relax
\mciteBstWouldAddEndPuncttrue
\mciteSetBstMidEndSepPunct{\mcitedefaultmidpunct}
{\mcitedefaultendpunct}{\mcitedefaultseppunct}\relax
\EndOfBibitem
\bibitem[Ceriotti \emph{et~al.}(2014)Ceriotti, More, and Manolopoulos]{ceriotti2014ipi}
M.~Ceriotti, J.~More and D.~E. Manolopoulos, \href{https://doi.org/10.1016/j.cpc.2013.10.027}{i-PI: A Python interface for ab initio path integral molecular dynamics simulations}, \emph{Computer Physics Communications}, 2014, \textbf{185}, 1019--1026, \href{https://doi.org/10.1016/j.cpc.2013.10.027}{DOI: 10.1016/j.cpc.2013.10.027}\relax
\mciteBstWouldAddEndPuncttrue
\mciteSetBstMidEndSepPunct{\mcitedefaultmidpunct}
{\mcitedefaultendpunct}{\mcitedefaultseppunct}\relax
\EndOfBibitem
\bibitem[Wang \emph{et~al.}(2018)Wang, Zhang, Han, and E]{wang2018deepmd}
H.~Wang, L.~Zhang, J.~Han and W.~E, \href{https://doi.org/10.1016/j.cpc.2018.03.016}{{DeePMD-kit}: A deep learning package for many-body potential energy representation and molecular dynamics}, \emph{Computer Physics Communications}, 2018, \textbf{228}, 178--184, \href{https://doi.org/10.1016/j.cpc.2018.03.016}{DOI: 10.1016/j.cpc.2018.03.016}\relax
\mciteBstWouldAddEndPuncttrue
\mciteSetBstMidEndSepPunct{\mcitedefaultmidpunct}
{\mcitedefaultendpunct}{\mcitedefaultseppunct}\relax
\EndOfBibitem
\bibitem[Batatia \emph{et~al.}(2022)Batatia, Kovacs, Simm, Ortner, and Csanyi]{Batatia2022mace}
I.~Batatia, D.~P. Kovacs, G.~N.~C. Simm, C.~Ortner and G.~Csanyi, \href{https://openreview.net/forum?id=YPpSngE-ZU}{{MACE}: Higher Order Equivariant Message Passing Neural Networks for Fast and Accurate Force Fields}, \href{https://openreview.net/forum?id=YPpSngE-ZU}{Advances in Neural Information Processing Systems}, 2022\relax
\mciteBstWouldAddEndPuncttrue
\mciteSetBstMidEndSepPunct{\mcitedefaultmidpunct}
{\mcitedefaultendpunct}{\mcitedefaultseppunct}\relax
\EndOfBibitem
\bibitem[Batatia \emph{et~al.}(2022)Batatia, Batzner, Kov{\'a}cs, Musaelian, Simm, Drautz, Ortner, Kozinsky, and Cs{\'a}nyi]{Batatia2022Design}
I.~Batatia, S.~Batzner, D.~P. Kov{\'a}cs, A.~Musaelian, G.~N.~C. Simm, R.~Drautz, C.~Ortner, B.~Kozinsky and G.~Cs{\'a}nyi, \emph{\href{https://doi.org/10.48550/arXiv.2205.06643}{The Design Space of E(3)-Equivariant Atom-Centered Interatomic Potentials}}, 2022, \url{https://doi.org/10.48550/arXiv.2205.06643}, \href{https://doi.org/10.48550/arXiv.2205.06643}{DOI: 10.48550/arXiv.2205.06643}\relax
\mciteBstWouldAddEndPuncttrue
\mciteSetBstMidEndSepPunct{\mcitedefaultmidpunct}
{\mcitedefaultendpunct}{\mcitedefaultseppunct}\relax
\EndOfBibitem
\bibitem[Gao(2023)]{deepmdjax2023}
R.~Gao, \emph{\href{https://github.com/SparkyTruck/deepmd-jax}{DeepMD-JAX: A JAX implementation of Deep Potential}}, 2023, \url{https://github.com/SparkyTruck/deepmd-jax}\relax
\mciteBstWouldAddEndPuncttrue
\mciteSetBstMidEndSepPunct{\mcitedefaultmidpunct}
{\mcitedefaultendpunct}{\mcitedefaultseppunct}\relax
\EndOfBibitem
\bibitem[Bigi \emph{et~al.}(2024)Bigi, Chong, Ceriotti, and Grasselli]{bigi2024rigidity}
F.~Bigi, S.~Chong, M.~Ceriotti and F.~Grasselli, \href{https://doi.org/10.1088/2632-2153/ad805f}{A prediction rigidity formalism for low-cost uncertainties in trained neural networks}, \emph{Machine Learning: Science and Technology}, 2024, \textbf{5}, 045018, \href{https://doi.org/10.1088/2632-2153/ad805f}{DOI: 10.1088/2632-2153/ad805f}\relax
\mciteBstWouldAddEndPuncttrue
\mciteSetBstMidEndSepPunct{\mcitedefaultmidpunct}
{\mcitedefaultendpunct}{\mcitedefaultseppunct}\relax
\EndOfBibitem
\bibitem[Chong \emph{et~al.}(2025)Chong, Bigi, Grasselli, Loche, Kellner, and Ceriotti]{chong2025rigidities}
S.~Chong, F.~Bigi, F.~Grasselli, P.~Loche, M.~Kellner and M.~Ceriotti, \href{https://doi.org/10.1039/D4FD00101J}{Prediction rigidities for data-driven chemistry}, \emph{Faraday Discussions}, 2025, \textbf{256}, 322--344, \href{https://doi.org/10.1039/D4FD00101J}{DOI: 10.1039/D4FD00101J}\relax
\mciteBstWouldAddEndPuncttrue
\mciteSetBstMidEndSepPunct{\mcitedefaultmidpunct}
{\mcitedefaultendpunct}{\mcitedefaultseppunct}\relax
\EndOfBibitem
\bibitem[Sun and Cheng(2019)]{sun2019phase}
J.-J. Sun and J.~Cheng, \href{https://doi.org/10.1038/s41467-019-13509-3}{Solid-to-liquid phase transitions of sub-nanometer clusters enhance chemical transformation}, \emph{Nature Communications}, 2019, \textbf{10}, 5400, \href{https://doi.org/10.1038/s41467-019-13509-3}{DOI: 10.1038/s41467-019-13509-3}\relax
\mciteBstWouldAddEndPuncttrue
\mciteSetBstMidEndSepPunct{\mcitedefaultmidpunct}
{\mcitedefaultendpunct}{\mcitedefaultseppunct}\relax
\EndOfBibitem
\bibitem[Fan \emph{et~al.}(2024)Fan, Gong, Liu, Zhu, and Cheng]{fan2024dynamic}
Q.-Y. Fan, F.-Q. Gong, Y.-P. Liu, H.-X. Zhu and J.~Cheng, \href{https://doi.org/10.1021/acscatal.4c05372}{Modeling Dynamic Catalysis at ab Initio Accuracy: The Need for Free-Energy Calculation}, \emph{ACS Catalysis}, 2024, \textbf{14}, 16086--16097, \href{https://doi.org/10.1021/acscatal.4c05372}{DOI: 10.1021/acscatal.4c05372}\relax
\mciteBstWouldAddEndPuncttrue
\mciteSetBstMidEndSepPunct{\mcitedefaultmidpunct}
{\mcitedefaultendpunct}{\mcitedefaultseppunct}\relax
\EndOfBibitem
\bibitem[Zhuang \emph{et~al.}(2025)Zhuang, Liu, Zhu, Hu, Le, Li, Wen, Fan, Jia, Li, Chen, Li, Lin, Xu, and Cheng]{zhuang_artificial_2025}
Y.-B. Zhuang, C.~Liu, J.-X. Zhu, J.-Y. Hu, J.-B. Le, J.-Q. Li, X.-J. Wen, X.-T. Fan, M.~Jia, X.-Y. Li, A.~Chen, L.~Li, Z.-L. Lin, W.-H. Xu and J.~Cheng, \href{https://doi.org/10.1038/s41597-025-05338-5}{An artificial intelligence accelerated ab initio molecular dynamics dataset for electrochemical interfaces}, \emph{Scientific Data}, 2025, \textbf{12}, 997, \href{https://doi.org/10.1038/s41597-025-05338-5}{DOI: 10.1038/s41597-025-05338-5}\relax
\mciteBstWouldAddEndPuncttrue
\mciteSetBstMidEndSepPunct{\mcitedefaultmidpunct}
{\mcitedefaultendpunct}{\mcitedefaultseppunct}\relax
\EndOfBibitem
\bibitem[Liu \emph{et~al.}(2025)Liu, Fan, Gong, and Cheng]{liu_catflow_2025}
Y.-P. Liu, Q.-Y. Fan, F.-Q. Gong and J.~Cheng, \href{https://doi.org/10.1021/acs.jpcc.4c05568}{{CatFlow}: An Automated Workflow for Training Machine Learning Potentials to Compute Free Energies in Dynamic Catalysis}, \emph{The Journal of Physical Chemistry C}, 2025, \textbf{129}, 1089--1102, \href{https://doi.org/10.1021/acs.jpcc.4c05568}{DOI: 10.1021/acs.jpcc.4c05568}\relax
\mciteBstWouldAddEndPuncttrue
\mciteSetBstMidEndSepPunct{\mcitedefaultmidpunct}
{\mcitedefaultendpunct}{\mcitedefaultseppunct}\relax
\EndOfBibitem
\bibitem[Shang and Liu(2013)]{shang_stochastic_2013}
C.~Shang and Z.-P. Liu, \href{https://doi.org/10.1021/ct301010b}{Stochastic Surface Walking Method for Structure Prediction and Pathway Searching}, \emph{Journal of Chemical Theory and Computation}, 2013, \textbf{9}, 1838--1845, \href{https://doi.org/10.1021/ct301010b}{DOI: 10.1021/ct301010b}\relax
\mciteBstWouldAddEndPuncttrue
\mciteSetBstMidEndSepPunct{\mcitedefaultmidpunct}
{\mcitedefaultendpunct}{\mcitedefaultseppunct}\relax
\EndOfBibitem
\bibitem[Shang \emph{et~al.}(2014)Shang, Zhang, and Liu]{shang_stochastic_2014}
C.~Shang, X.-J. Zhang and Z.-P. Liu, \href{https://doi.org/10.1039/C4CP01485E}{Stochastic surface walking method for crystal structure and phase transition pathway prediction}, \emph{Phys. Chem. Chem. Phys.}, 2014, \textbf{16}, 17845--17856, \href{https://doi.org/10.1039/C4CP01485E}{DOI: 10.1039/C4CP01485E}\relax
\mciteBstWouldAddEndPuncttrue
\mciteSetBstMidEndSepPunct{\mcitedefaultmidpunct}
{\mcitedefaultendpunct}{\mcitedefaultseppunct}\relax
\EndOfBibitem
\bibitem[Guo \emph{et~al.}(2023)Guo, Zhuang, Shi, and Cheng]{guo_checmate_2023}
Y.-X. Guo, Y.-B. Zhuang, J.~Shi and J.~Cheng, \href{https://doi.org/10.1063/5.0166858}{{ChecMatE}: A workflow package to automatically generate machine learning potentials and phase diagrams for semiconductor alloys}, \emph{The Journal of Chemical Physics}, 2023, \textbf{159}, 094801, \href{https://doi.org/10.1063/5.0166858}{DOI: 10.1063/5.0166858}\relax
\mciteBstWouldAddEndPuncttrue
\mciteSetBstMidEndSepPunct{\mcitedefaultmidpunct}
{\mcitedefaultendpunct}{\mcitedefaultseppunct}\relax
\EndOfBibitem
\bibitem[Wang and Cheng(2022)]{wang2022automated}
F.~Wang and J.~Cheng, \href{https://doi.org/10.1063/5.0098330}{Automated workflow for computation of redox potentials, acidity constants, and solvation free energies accelerated by machine learning}, \emph{The Journal of Chemical Physics}, 2022, \textbf{157}, 024103, \href{https://doi.org/10.1063/5.0098330}{DOI: 10.1063/5.0098330}\relax
\mciteBstWouldAddEndPuncttrue
\mciteSetBstMidEndSepPunct{\mcitedefaultmidpunct}
{\mcitedefaultendpunct}{\mcitedefaultseppunct}\relax
\EndOfBibitem
\bibitem[Zou(2026)]{mokit2026}
J.~Zou, \emph{\href{https://gitlab.com/jxzou/mokit}{Molecular {Orbital} {Kit} ({MOKIT})}}, 2026, \url{https://gitlab.com/jxzou/mokit}\relax
\mciteBstWouldAddEndPuncttrue
\mciteSetBstMidEndSepPunct{\mcitedefaultmidpunct}
{\mcitedefaultendpunct}{\mcitedefaultseppunct}\relax
\EndOfBibitem
\bibitem[Zhang \emph{et~al.}(2020)Zhang, Chen, Wu, Wang, E, and Car]{zhang2020deepwannier}
L.~Zhang, M.~Chen, X.~Wu, H.~Wang, W.~E and R.~Car, \href{https://doi.org/10.1103/PhysRevB.102.041121}{Deep neural network for the dielectric response of insulators}, \emph{Physical Review B}, 2020, \textbf{102}, 041121, \href{https://doi.org/10.1103/PhysRevB.102.041121}{DOI: 10.1103/PhysRevB.102.041121}\relax
\mciteBstWouldAddEndPuncttrue
\mciteSetBstMidEndSepPunct{\mcitedefaultmidpunct}
{\mcitedefaultendpunct}{\mcitedefaultseppunct}\relax
\EndOfBibitem
\bibitem[Ahrens-Iwers \emph{et~al.}(2022)Ahrens-Iwers, Janssen, Tee, and Mei{\ss}ner]{ahrens2022electrode}
L.~J.~V. Ahrens-Iwers, M.~Janssen, S.~R. Tee and R.~H. Mei{\ss}ner, \href{https://doi.org/10.1063/5.0099239}{{ELECTRODE}: An electrochemistry package for atomistic simulations}, \emph{The Journal of Chemical Physics}, 2022, \textbf{157}, 084801, \href{https://doi.org/10.1063/5.0099239}{DOI: 10.1063/5.0099239}\relax
\mciteBstWouldAddEndPuncttrue
\mciteSetBstMidEndSepPunct{\mcitedefaultmidpunct}
{\mcitedefaultendpunct}{\mcitedefaultseppunct}\relax
\EndOfBibitem
\bibitem[Zhu and Cheng(2025)]{ec-MLP}
J.-X. Zhu and J.~Cheng, \href{https://doi.org/10.1103/48ct-3jxm}{Machine Learning Potential for Electrochemical Interfaces with Hybrid Representation of Dielectric Response}, \emph{Physical Review Letters}, 2025, \textbf{135}, 018003, \href{https://doi.org/10.1103/48ct-3jxm}{DOI: 10.1103/48ct-3jxm}\relax
\mciteBstWouldAddEndPuncttrue
\mciteSetBstMidEndSepPunct{\mcitedefaultmidpunct}
{\mcitedefaultendpunct}{\mcitedefaultseppunct}\relax
\EndOfBibitem
\bibitem[Sommers \emph{et~al.}(2020)Sommers, Calegari~Andrade, Zhang, Wang, and Car]{sommers2020raman}
G.~M. Sommers, M.~F. Calegari~Andrade, L.~Zhang, H.~Wang and R.~Car, \href{https://doi.org/10.1039/D0CP01893G}{Raman spectrum and polarizability of liquid water from deep neural networks}, \emph{Physical Chemistry Chemical Physics}, 2020, \textbf{22}, 10592--10602, \href{https://doi.org/10.1039/D0CP01893G}{DOI: 10.1039/D0CP01893G}\relax
\mciteBstWouldAddEndPuncttrue
\mciteSetBstMidEndSepPunct{\mcitedefaultmidpunct}
{\mcitedefaultendpunct}{\mcitedefaultseppunct}\relax
\EndOfBibitem
\bibitem[Larouche \emph{et~al.}(2008)Larouche, Max, and Chapados]{larouche2008ir}
P.~Larouche, J.-J. Max and C.~Chapados, \href{https://doi.org/10.1063/1.2960583}{Isotope Effects in Liquid Water by Infrared Spectroscopy. II. Factor Analysis of the Temperature Effect on H2O and D2O}, \emph{The Journal of Chemical Physics}, 2008, \textbf{129}, 064503, \href{https://doi.org/10.1063/1.2960583}{DOI: 10.1063/1.2960583}\relax
\mciteBstWouldAddEndPuncttrue
\mciteSetBstMidEndSepPunct{\mcitedefaultmidpunct}
{\mcitedefaultendpunct}{\mcitedefaultseppunct}\relax
\EndOfBibitem
\bibitem[Scherer \emph{et~al.}(1974)Scherer, Go, and Kint]{scherer1974raman}
J.~R. Scherer, M.~K. Go and S.~Kint, \href{https://doi.org/10.1021/j100606a013}{Raman Spectra and Structure of Water from -10 to 90.deg.}, \emph{The Journal of Physical Chemistry}, 1974, \textbf{78}, 1304--1313, \href{https://doi.org/10.1021/j100606a013}{DOI: 10.1021/j100606a013}\relax
\mciteBstWouldAddEndPuncttrue
\mciteSetBstMidEndSepPunct{\mcitedefaultmidpunct}
{\mcitedefaultendpunct}{\mcitedefaultseppunct}\relax
\EndOfBibitem
\bibitem[Wang \emph{et~al.}(2024)Wang, Tang, Yu, Ohto, Nagata, and Bonn]{wang2024sfg}
Y.~Wang, F.~Tang, X.~Yu, T.~Ohto, Y.~Nagata and M.~Bonn, \href{https://doi.org/10.1002/anie.202319503}{Heterodyne-Detected Sum-Frequency Generation Vibrational Spectroscopy Reveals Aqueous Molecular Structure at the Suspended Graphene/Water Interface}, \emph{Angewandte Chemie International Edition}, 2024, \textbf{63}, e202319503, \href{https://doi.org/10.1002/anie.202319503}{DOI: 10.1002/anie.202319503}\relax
\mciteBstWouldAddEndPuncttrue
\mciteSetBstMidEndSepPunct{\mcitedefaultmidpunct}
{\mcitedefaultendpunct}{\mcitedefaultseppunct}\relax
\EndOfBibitem
\end{mcitethebibliography}

\end{document}